%% file: Vicard.tex
\documentclass[sn-mathphys,Numbered]{sn-jnl}

\usepackage{graphicx}%
\usepackage{multirow}%
\usepackage{amsmath,amssymb,amsfonts}%
\usepackage{amsthm}%
\usepackage{mathrsfs}%
\usepackage[title]{appendix}%
\usepackage{xcolor}%
\usepackage{textcomp}%
\usepackage{manyfoot}%
\usepackage{booktabs}%
\usepackage{algorithm}%
\usepackage{algorithmicx}%
\usepackage{algpseudocode}%
\usepackage{listings}%

\usepackage{lineno}
\usepackage[utf8]{inputenc}
\usepackage[english]{babel}
\usepackage{amsmath}
\usepackage{amsfonts}
\usepackage{amssymb}
\usepackage{graphicx}
\usepackage{bm}
\usepackage{xcolor}
\usepackage{natbib}
\usepackage{float}
\usepackage{subfig}
\usepackage[font=footnotesize]{caption}
\usepackage{setspace}

\usepackage{multicol}
\usepackage{afterpage}

\newtheorem{Lemma}{Lemma}
\newtheorem{prop}{Proposition}

\newcommand\independent{\protect\mathpalette{\protect\independenT}{\perp}}
\def\independenT#1#2{\mathrel{\rlap{$#1#2$}\mkern2mu{#1#2}}}

\def\independenT#1#2{\mathrel{\rlap{$#1#2$}\mkern2mu{#1#2}}}

\begin{document}

\title[Article Title]{Testing for causal effect for binary data when propensity scores are estimated through Bayesian Networks}



\author[1]{\fnm{Paola} \sur{Vicard}}\email{paola.vicard@uniroma3.it}

\author[2]{\fnm{Paola Maria Vittoria} \sur{Rancoita}}\email{rancoita.paolamaria@unisr.it}

\author[2]{\fnm{Federica} \sur{Cugnata}}\email{cugnata.federica@unisr.it}

\author[2]{\fnm{Alberto} \sur{Briganti}}\email{briganti.alberto@hsr.it}
\author[3]{\fnm{Fulvia} \sur{Mecatti}}\email{fulvia.mecatti@unimib.it}
\author[2]{\fnm{Clelia} \sur{ Di Serio}}\email{clelia.diserio@unisr.it}
\author*[4]{\fnm{Pier Luigi} \sur{Conti}}\email{pierluigi.conti@uniroma1.it}

\affil[1]{\orgname{Roma Tre University}, \city{Rome}, \country{Italy}}
\affil[2]{\orgname{Vita-Salute San Raffaele University}, \city{Milan}, \country{Italy}}
\affil[3]{\orgname{University of Milano-Bicocca}, \city{Milan}, \country{Italy}}
\affil[4]{\orgname{Sapienza University of Rome}, \city{Rome}, \country{Italy}}


\abstract{This paper proposes a new statistical approach for assessing treatment effect using Bayesian Networks (BNs). The goal  is to draw causal inferences from observational data with a binary outcome and discrete covariates. The BNs are here used to estimate the propensity score, which enables flexible modeling and ensures maximum likelihood properties, including asymptotic efficiency. 
When the propensity score is estimated by BNs, two point estimators are considered-  H\'{a}jek
and  Horvitz-Thompson -  based on inverse probability weighting, and their main distributional properties are derived for constructing confidence intervals and testing hypotheses about the absence of the treatment effect.
Empirical evidence is presented to show the goodness of the proposed methodology on a simulation study mimicking the characteristics of a real dataset of prostate cancer patients from Milan San Raffaele Hospital.
}



\maketitle

\section{Introduction}\label{sec1}
\label{introd}
Observational data in biomedical research are progressively gaining significance, owing to increasing costs and regulatory hurdles associated with conducting clinical trials.
However, its potential for external validity is still being debated.
Indeed, working with observational data has many challenges, such as non-randomized (uncontrolled) assignment of treatments and non-controlled study designs.
Developing new rigorous statistical methods for overcoming these problems could potentially improve generalizability.

From a statistical perspective, the uncontrolled unbalanced design often associated to observational data could be approached by  estimating the probability of receiving the treatments given the observed covariates (propensity score, PS). The PS can be then used for rebalancing the sample while statistically analyzing the data. The actual statistical approaches for evaluating the treatment effects in this context are mainly based on: $(i)$ matching (possibly based on PS); $(ii)$ stratification matching (possibly based on PS); and $(iii)$ inverse probability weighting (IPW) through estimated PS.
In this article, we will focus on the latter approach, by defining a new test for the absence of treatment effect within the inverse probability weighting approach, as the other methods, even when using PS, have various disadvantages.
For example, while stratification matching is useful for estimating the Average Treatment Effect (ATE), it is problematic to use for testing the presence of treatment effects and does not allow for straightforward estimation beyond the ATE point. In contrast, IPW avoids these limitations by easily extending to identifiable parameters other than ATE and enabling the construction of tests for treatment effect (cfr. \cite{condg22}, \cite{donhsu14} and references therein). Additionally, stratification matching requires a careful choice of the number of strata, which is subjective and must increase with the sample size to ensure consistency of estimates. Unfortunately, it is unclear how fast the number of strata should increase, which can lead to theoretically unsafe estimates. Furthermore, when stratification is based on covariate values, the number of resulting strata can become too numerous as the number of covariates and/or their levels increase, leading to insufficient data issues due to excess of granularity.
Matching estimators for treatment effects are mainly studied under the assumption of continuous covariates with positive density; (cfr. the fundamental paper by \cite{abadieImbens:16} and references therein). The extension to discrete outcomes, as in the present paper, requires non-trivial theoretical changes and is beyond the scope of this paper. Moreover, matching estimators do not easily include parameters other than ATE, and their use as a test statistic is problematic.

A common feature of the above-mentioned methods (matching, stratification matching, and IPW) is that they typically require the estimation of PS.
In this paper, we propose using Bayesian Networks (BNs) for PS estimation. BNs can capture the dependence structure between covariates and the treatment assignment in a data-driven perspective. By their nature, they have a clear interpretation in terms of both dependencies among covariates and the influence of covariates on the probability of receiving treatment.
Therefore, BNs are flexible 
and they also have the key advantage of providing consistent and asymptotically efficient estimates of propensity scores under the setting of the present paper (categorical covariates). This is the strongest theoretical argument in favor of their use.

In conclusion, this paper has two main objectives:
\begin{itemize}
\item[1.] to propose a new Bayesian Networks-based estimator for propensity score, with sound statistical properties;
\item[2.] to construct a test-statistic for the evaluation of the null hypothesis of the absence of treatment effect, by estimating potential outcome probabilities based on the estimated propensity score.
\end{itemize}
The paper is organized as follows.
In Section \ref{sec:basic_theory}, we introduce the fundamental concepts related to propensity score and Bayesian Networks. In Section \ref{sec:theory}, we explore the properties of the Bayesian Networks-based propensity score estimators. Additionally, we examine two types of ATE estimator in the IPW class using the obtained propensity score. If the PS model is correctly specified, they are asymptotically equivalent, and their statistical efficiency is the same for large datasets.
Then, we develop a statistical test to detect the presence of the treatment effect. In Section \ref{sec:misspecification}, we show that, under PS specification error, the two types of ATE estimators behave differently and have varying degrees of robustness with respect to misspecification.
Finally, in Section \ref{sec:mot-ex}, we present a motivating case study based on a large observational real data from prostate cancer patients. Section \ref{sec:empir-RWD} describes a simulation study evaluating the performance of the considered ATE estimators and the corresponding tests. 

\section{Basic theory}
\label{sec:basic_theory}

\subsection{Notation and basic concepts}
\label{sec:notation}

Consider a random sample of $n$ independent subjects. Each of them could either receive or not a treatment $T$; conventionally,  $T=1$
means a subject does receive the treatment, whilst $T=0$ means the subject does not receive the treatment.
Let  $Y_{(1)}$ denote the potential outcome of a subject in the presence of the treatment ({\em i.e.} when $T=1$), and let
$Y_{(0)}$ denote the potential outcome of a subject in the absence of the treatment ({\em i.e.} when $T=0$).
The observed outcome for a subject is then:
\begin{eqnarray}
Y = T Y_{(1)} + (1-T) Y_{(0)}. \label{eq:obs-outc-1}
\end{eqnarray}

Eqn. (\ref{eq:obs-outc-1}) can be more conveniently written by using the indicator function of the treatment.
Let $I_{(T = 1)}$ be the indicator of the event $T=1$, namely
\begin{eqnarray}
I_{(T = 1)} = \left \{
\begin{matrix}
& 1 &   {\mathrm{if}} \; T=1 \\
& 0 &   {\mathrm{if}} \; T=0,
\end{matrix}
\right . 
\nonumber
\end{eqnarray}
\noindent thus
 $I_{(T = 0)} = 1- I_{(T = 1)}$ is the indicator function of the event $T=0$.
The observed outcome $Y$ can be then written as
\begin{eqnarray}
Y =  Y_{(1)} I_{(T = 1)} + Y_{(0)} I_{(T = 0)}. \label{eq:observed_outc}
\end{eqnarray}

The treatment has no effect when $Y_{(0)}$ and $Y_{(1)}$ have the same probability distribution. If $\stackrel{d}{=}$ denotes
equality in distribution, then there is absence of treatment if and only if (iff) $Y_{(0)} \stackrel{d}{=} Y_{(1)}$.
This is, in a sense, a ``probabilistic version'' for observational data of
the {\em sharp hypothesis} of absence of treatment effect in experimental studies; cfr. \cite{ding17}, \cite{wuding18}.

The assignment-to-treatment mechanism is not a ``purely random" mechanism,
as in experimental framework. There could be considerable differences among subjects receiving different treatment levels due to the presence of confounding covariates.
As usual in the literature (cfr. \cite{imbrub:15}), we assume here that the assignment-to-treatment mechanism only depends on
a set of $L$ observed covariates; from now on,
the vector of relevant covariates is denoted by $\bm{X} = ( X_1 \, \cdots \, X_L )$. Furthermore, the probability of receiving the treatment
 conditionally on $\bm{X} = \bm{x}$, namely the {\em propensity score}, is denoted by
\begin{eqnarray}
p_1 ( \bm{x} ) = P ( T=1 \vert \bm{X} = \bm{x} ) . \label{eq:prop_sc}
\end{eqnarray}
For the sake of completeness, in the sequel we will denote by
\begin{eqnarray}
p_0 ( \bm{x} ) = P ( T=0 \vert \bm{X} = \bm{x} )
\end{eqnarray}
\noindent the probability of not receiving the treatment, given $\bm{X} = \bm{x}$.
The relationship $p_0 ( \bm{x} ) + p_1 ( \bm{x} ) = 1$ holds.

In the present paper, we focus on the case where covariates  are
discrete finite random variables (r.v.s), and the potential outcomes are dichotomic variables.
The assumptions on which our analysis rests are listed below.

\begin{itemize}
\item[A1.] {\em Discreteness}. Each covariate $X_l$ is discrete, finite. With no loss of generality, it may take nominal values
$1, \, \dots , \, r_l$ with
positive probability. The potential outcomes $Y_{(k)}$ are dichotomic r.v.s. For the sake of
simplicity, and with no loss of generality, in the sequel we assume that $Y_{(k)}$, $k=0, \, 1$, may take values $0$, $1$ with (marginal) probability:
\begin{eqnarray}
\theta_k  = P ( Y_{(k)} = 1 ) ; \;\; k= 0, \, 1 .
\label{eq:prob_theta}
\end{eqnarray}
\item[A2.] {\em Unconfoundedness}. $T \independent (Y_{(0)} , \, Y_{(1)}  ) \vert  \bm{X}$, where the symbol $\independent $
denotes stochastic independence.
\item[A3.] {\em Common support}. There exists a positive real $\delta$ for which $ \delta \leq p_k ( \bm{x} ) \leq 1- \delta$ for each
$\bm{x}$ and $k=0, \, 1$.
\end{itemize}

In the case under examination, the absence of treatment effect is equivalent to say that $\theta_0  = \theta_1 $.
Hence, testing for the absence of treatment effect reduces to the following hypothesis problem

\begin{eqnarray}
\left \{
\begin{matrix}
& H_0 : &   \theta_1 =  \theta_0  \\
& H_1 : &   \theta_1  \neq  \theta_0
\end{matrix}
\right .
\label{eq:hp_absence_treatm}
\end{eqnarray}

The hypothesis problem (\ref{eq:hp_absence_treatm}) can be also expressed in terms of the familiar notion of Average Treatment Effect (ATE), defined as
$ATE = E[ Y_{(1)} ] - E[ Y_{(0)} ]$. Since, from definition (\ref{eq:prob_theta}),
 $E[ Y_{(1)} ] = \theta_1$, $E[ Y_{(0)} ] = \theta_0 $, the null hypothesis is
equivalent to $ATE =0$.

Observed data for $n$ subjects are formally defined as the triplets $( Y_i , \, T_i , \, \bm{X}_i )$, $i=1, \, \dots , \, n$. The r.v.s $( Y_i , \, T_i , \, \bm{X}_i )$
are assumed to be independent and identically distributed ({\em i.i.d.}).

As it appears from  \cite{imbenswool09}, testing whether the distribution of
$Y_{(0)}$ is different from that of $Y_{(1)}$ is a problem of considerable interest, and relevance, as well.
Bootstrap-based tests in the settings with randomized experiments are studied in \cite{abadie02}; again in the setting of randomized experiments,
permutation tests are studied in \cite{ding17}, \cite{wuding18}.
Specifically devoted to the case of ordinal qualitative outcomes is the paper \cite{luzhangding19}, where a test for an upper bound for
$\gamma = P( Y_{(1)} > Y_{(0)} ) -  P( Y_{(1)} < Y_{(0)} )$ is proposed; for further articles dealing with this subject, the reader is deferred to
references in the above mentioned paper.

\subsection{Bayesian Networks: basic aspects}
\label{sec:BN}

The basic idea exploited in the present section is to estimate propensity scores $p_k ( \bm{x} )$ by using a Bayesian Network  model for
$(T, \, \bm{X} )
$; cfr. \cite{cowell99}.

To simplify the notation, the r.v. $T$ is here denoted by $X_0$, so that $(T, \, \bm{X} ) = ( X_0 , \, X_1 , \, \dots , \,
X_L ) = ( X_0 , \, \bm{X} )$.
A BN is a multivariate statistical model satisfying sets of conditional independence statements displayed in a directed acyclic graph (DAG), consisting in a set of {\em nodes} and a set of {\em directed arcs} connecting pairs of nodes. The nodes represent random variables and the arcs represent direct dependencies among the variables. Each node is associated with an index $l =0, \, 1, \, \dots , L$, which, in its turn, corresponds to a the random variable $X_l$.
A directed graph is acyclic in the sense that it is forbidden to start
from a node and, following arrows directions, go back to the starting node.

Next, let 
$pa(l)$ be the set of {\em parent nodes} of node $l$, {\em i.e.} the set of all nodes with a directed arc pointing to
node $l$.  In equivalent terms, $\bm{X}_{pa(l)}$ is the set of all variables in $(X_0, \, \bm{X} )$
``graphically'' linked to $X_l$.
Conditional independencies can be read off the DAG by means of the Markov properties.
Specifically, the joint probability distribution over a DAG satisfies the local Markov property if any variable, say $X_l$, is
independent of its non-descendants conditionally on its parents, where the set of non-descendents of $X_l$ is composed by
all the variables for which there is no directed path from $X_l$ to them. For other Markov properties for DAGs (that have been shown to be equivalent), cfr. \cite{lauritzen96}.
In a BN, each node of a DAG is associated with the distribution of the corresponding variable given its parents (if a node has no parents, it is associated with its marginal distribution). Formally speaking a BN is a pair DAG/joint probability distribution satisfying the Markov properties. 
The joint distribution can be factorised according to the DAG as the product of the conditional distributions associated to each node given its parents.
In general, the chain rule for the distribution of $(X_0, X_1,\dots, X_L)$ states that:
\begin{eqnarray}
p ( X_0 = x_0 , \, \bm{X} =  \bm{x} ) =  p( x_0 , \, x_1 , \, \dots , \, x_L ) = \prod_{l=0}^{L}  p( x_l \vert \bm{x}_{pa(l)} ) .
\nonumber
\end{eqnarray}

When covariates $\mathbf{X}$ are discrete, the use of BNs to estimate propensity scores has several positive theoretical features and practical advantages. 
On the
theoretical side, BNs allow to consider Maximum Likelihood Estimators (MLEs) of propensity scores $p_k ( \bm{x} )$, that possess ``usual'' properties
of MLEs, {\em  i.e.}  they are $\sqrt{n}$-consistent and asymptotically efficient. Computation of MLEs in BNs is carefully studied in \cite{cowell99}.
More explicitly, for each $l$ denote by $\mathbf{P}_l$ the Cartesian product
\begin{eqnarray}
\mathbf{P}_l = \prod_{j \in pa(l)} \{ 1, \, \dots , \, r_j \} .
\nonumber
\end{eqnarray}
\emph{i.e.} the parent set configuration of node $l$, $l =0, \, 1, \, \dots , L$.
The parameters of the BN are the conditional probabilities
\begin{eqnarray}
\lambda_{x_l \vert \bm{x}_{pa(l)}} =  p( x_l \vert \bm{x}_{pa(l)} ) , \;\; x_l \in \{ 1, \, \dots , \, r_l \} , \;
\bm{x}_{pa(l)} \in \mathbf{P}_l , \;
l=0, \, 1 , \, \dots , \, L .
\label{eq:par-bn}
\end{eqnarray}

The constraints
\begin{eqnarray}
\sum_{x_{l} =1}^{r_{l}} \lambda_{x_l \vert \bm{x}_{pa(l)}}  = 1 \;\; \forall \, \bm{x}_{pa(l)} \in \mathbf{P}_l , \;
l=0, \, 1, \, \dots L
\nonumber
\end{eqnarray}
\noindent hold.

Let $ \bm{\lambda}$ be the vector of parameters (\ref{eq:par-bn}). The likelihood function for the BN under consideration essentially corresponds to
a Multinomial likelihood with parameters vector $ \bm{\lambda}$.
If $x_{il}$ denotes the value of the variable $X_l$ for the $i$th observation, and $ \bm{x}_{i pa(l)} $ are the values of the parents of $X_l$ for
the $i$th observation, the log-likelihood
is equal to
\begin{eqnarray}
l ( \bm{\lambda} ) & = & \sum_{i=1}^{n} \sum_{l=0}^{L} \log \lambda_{x_{il} \vert \bm{x}_{i pa(l)}} \nonumber \\
& = & \sum_{l=0}^{L} \sum_{\bm{x}_{pa(l)} \in \mathbf{P}_l} \sum_{x_{l} =1}^{r_{l}}
n \left ( x_{l}, \, \bm{x}_{pa(l)} \right ) \log  \lambda_{x_{l} \vert \bm{x}_{pa(l)}}  \nonumber
\end{eqnarray}
\noindent where $n \left ( x_{l}, \, \bm{x}_{pa(l)} \right )$ is the number of times the pair $\left ( x_l , \,  \bm{x}_{pa(l)} \right )$ is observed in the sample.
Hence, the MLE of $\lambda_{x_l \vert \bm{x}_{pa(l)}} $ is just the empirical conditional proportion
\begin{eqnarray}
\widehat{\lambda}_{x_l \vert \bm{x}_{pa(l)}} = \frac{n \left ( x_{l}, \, \bm{x}_{pa(l)} \right )
}{n \left ( \bm{x}_{pa(l)} \right )}
\nonumber
\end{eqnarray}
\noindent where
\begin{eqnarray}
n \left ( \bm{x}_{pa(l)} \right ) = \sum_{x_{l} =1}^{r_{l}} n \left ( x_{l}, \, \bm{x}_{pa(l)} \right ) \nonumber
\end{eqnarray}
\noindent is the number of times $pa(x_{l} )$ is observed in the sample.

The MLE of $p( x_0 , \, x_1 , \, \dots , \, x_L )$  is obtained by the invariance property:
\begin{eqnarray}
\widehat{p} ( x_0 , \, x_1 , \, \dots , \, x_L ) = \prod_{l=0}^{L} \widehat{\lambda}_{x_l \vert \bm{x}_{pa(l)}} =
\prod_{l=0}^{L} \frac{n \left ( x_{l}, \, \bm{x}_{pa(l)} \right )
}{n \left ( \bm{x}_{pa(l)} \right )}.
\label{eq:MLE_joint}
\end{eqnarray}

The association structure, i.e. the presence or absence of edges and their direction, between the variables and their conditional probability distributions can be either known in advance by subject-matter knowledge or has to be learnt (estimated) from the data. In the second case, it is crucial to appropriately estimate the dependence model. To this aim there are three main classes of learning algorithms: constraint-based, score-based and hybrid algorithms (cfr. \cite{drtmal17}). In constraint-based learning, the DAG is learned according to a sequence of independence tests carried out on data, and to logic rules based on test results.
In score-based learning, it is necessary to compute the maximum likelihood of each structure, penalize it in order to avoid overfitting (the penalized likelihood is also named score), and search for the structure with the best score. Penalization is usually performed based on AIC or BIC. When the number of nodes is greater than 5, the number of possible network structures is so large that it is necessary to resort to appropriate search algorithms such as the greedy-search algorithm. Finally, hybrid algorithms combine the ideas and the
advantages of the previous two types of algorithms.
Compared to traditional statistical models, BNs allow a high degree of flexibility in discovering the dependence relationships among covariates and treatment levels, that is the presence of possible relationships of conditional independence (i.e. for the presence/absence of arcs) in the model, without requiring the identification, for each treatment level, of polynomial terms and interactions terms for the covariates in $\bm{x}$ to be included in the model.

In the sequel, theoretical results will be illustrated with a three-fold purpose. First of all, we consider estimates of propensity scores
on the basis of an appropriate Bayesian Network (BN) model. As it will be seen, the obtained estimates possess highly desirable statistical properties. In the second place, two types of ATE estimators based on propensity score weighting are considered, and their statistical properties are studied. As it will be also seen, one of them is usually better 
under misspecification of the propensity score model. 
Finally, estimated propensity scores are used to construct a test statistic for the presence of treatment effect based on an appropriate divergence measure.

\section{Theoretical results}
\label{sec:theory}

\subsection{Estimation of propensity scores by Bayesian Network models}
\label{sec:estim_prop_score}

As already said, due to their flexibility in modeling the dependence structures among covariates, BNs naturally offer an excellent tool to estimate propensity scores.
From (\ref{eq:MLE_joint}), and using again the invariance property, the MLE of the propensity score $p_k ( \bm{x} )$ is equal to
\begin{eqnarray}
\widehat{p}_k ( \bm{x} ) = \frac{\widehat{p} ( k , \, x_1 , \, \dots , \, x_L )}{\widehat{p} ( 1 , \, x_1 , \, \dots , \, x_L )
+ \widehat{p} ( 0 , \, x_1 , \, \dots , \, x_L )} , \;\; k=0, \, 1. \label{eq:MLE_propscore_BN}
\end{eqnarray}

The theoretical properties of (\ref{eq:MLE_propscore_BN}) follow as a  consequence of general results on MLEs (cfr. Davison (2003), \cite{lecam53}, \cite{lecam60}), and the same also holds for the marginal probability distribution of $( x_1 , \, \dots , \, x_L )$.

In order to establish results in the proper way, in the sequel we will denote by $p^{TRUE}_k ( \bm{x} )$ the
{\em true} propensity scores, and by  $\mathcal{P} = \{ ( p_0 ( \bm{x} ; \, \bm{\lambda}) ) ,  \, p_1 ( \bm{x} ; \; \bm{\lambda}) ;
\; \bm{\lambda} \in \bm{\Lambda} \}$ the {\em working} statistical model for propensity scores under BNs.
As an immediate consequence of the above mentioned general results, the following results for the propensity score estimated through BNs are obtained.

\begin{prop}
\label{prop_MLE}
Suppose that assumptions $A1$-$A3$ are met.
The following  results hold.
\begin{itemize}
\item[$(a)$] $ p^{TRUE}_k ( \bm{x} ) \in \mathcal{P}$.
\item[$(b)$] The estimator (\ref{eq:MLE_propscore_BN}) is consistent:
\begin{eqnarray}
\widehat{p}_k ( \bm{x} )   \stackrel{p}{\rightarrow} p_k ( \bm{x} )   \;\;
{\mathrm{as}} \; n \rightarrow \infty  , \;\; k=0, \, 1
\label{eq:cons_pk_BN}
\end{eqnarray}
\noindent for every value of $\bm{x}$, the symbol $\stackrel{p}{\rightarrow}$
denoting convergence in probability.
\item[$(c)$] The estimator (\ref{eq:MLE_propscore_BN}) is asymptotically Normally distributed:
\begin{eqnarray}
\sqrt{n} ( \widehat{p}_k ( \bm{x} )   -  p_k ( \bm{x} )  ) \stackrel{d}{\rightarrow}
N(0, \, I_k^{-1}( \bm{x} ) ) \;\;
{\mathrm{as}} \; n \rightarrow \infty  , \;\; k=0, \, 1
\label{eq:asympt-norm_pk_BN}
\end{eqnarray}
\noindent where  $I_k^{-1}( \bm{x} ) $ is the reciprocal of the Fisher information, and $\stackrel{d}{\rightarrow}$
denotes convergence in distribution.
\item[$(d)$] The estimator (\ref{eq:MLE_propscore_BN}) is asymptotically efficient. If $\widetilde{p}_k ( \bm{x} )$ is another (sequence of) estimator(s) of the PS $p_k ( \bm{x} )$, consistent and
asymptotically Normal with $\sqrt{n} ( \widetilde{p}_k ( \bm{x} )   -  p_k ( \bm{x} )  ) \stackrel{d}{\rightarrow}
N(0, \, V_k ( \bm{x} ) )$ as $n \rightarrow \infty$, then:
\begin{eqnarray}
V_k ( \bm{x} ) \geq I_k^{-1}( \bm{x} )
\label{eq:asympt-efficient_pk_BN}
\end{eqnarray}
\noindent for all $\bm{x}$s and $k=0, \, 1$.
\end{itemize}
\end{prop}

Notice that Proposition \ref{prop_MLE} is ``universal'', in the sense that it holds true whatever the ``true'' value of the propensity score may be. In other terms, statement $(a)$ above ensures that the true PS is always a member of the working
statistical model for BNs, thus avoiding the possibility of misspecification errors asymptotically (under conditions A1-A3, of course), even when their probabilistic structure is learnt from data. 

Proposition \ref{prop_MLE} actually offers the most relevant theoretical support to estimate propensity score under the conditions
A1-A3 of the present paper. In fact, it shows that the estimator (\ref{eq:MLE_propscore_BN}) obtained through BNs is not only consistent and asymptotically normally distributed, but it is also {\em asymptotically efficient}.
Competitor PS estimators can be either as  efficient as those obtained through BNs, or they are beated in terms of asymptotic MSE, which is a reason in favour of using BNs.

One of the nicest features of BNs is the possibility to learn their probabilistic structure from data.
From  \cite{balov13}, \cite{nandy18}, it is possible to argue that even in this case the above convergence results still hold.

We note {\em in passim} that, since \cite{hahn98}, the estimation of propensity score is usually considered as ancillary in estimating the parameter $\theta_1 - \theta_0$. Accordingly, a different approach to weighting consists in using moment method based on appropriate balancing with respect to covariates, or better in controlling the maximal imbalance; cfr. \cite{zubi15}, \cite{chatto20}, \cite{benmi21}. However, the speed of convergence of $\widehat{p}_k ( \bm{x} )$ has a non-trivial consequence in terms of imbalance. According to \cite{benmi21},
consider a bounded function $f ( \bm{x} )$, and let the imbalance with respect to $f$ be equal to
\begin{eqnarray}
\left \vert \frac{1}{n} \sum_{i=1}^{n} I_{(T_i =k)} \widehat{p}_k ( \bm{x}_i )^{-1} f( \bm{x}_i ) - \frac{1}{n} \sum_{i=1}^{n}
f( \bm{x}_i ) \right \vert . \label{eq:imbal1}
\end{eqnarray}
The Law of Large Numbers and the consistency of $\widehat{p}_k$ guarantee that  (\ref{eq:imbal1}) tends in probability to $0$ as $n$
increases. Furthermore, using Lemma \ref{lemma_1} (see Appendix) it is easy to see that (\ref{eq:imbal1}) tends to $0$ in probability at the same speed at which $\sup_{\bm{x}} \left \vert \widehat{p}_k ( \bm{x}_i )^{-1} - p_k ( \bm{x} )^{-1} \right \vert$ tends to $0$. In our case, due to
the assumptions made, the above supremum is $O_P ( n^{-1/2} )$, which is the maximal speed. In other words, even if the estimation of propensity score is ancillary for $\theta_1 - \theta_0$, it is of importance in terms of imbalance.

\subsection{Estimation of potential outcomes probabilities}
\label{sec:estim_pot_outc_prob}

In view of the estimator (\ref{eq:MLE_propscore_BN}) of propensity score,
an estimator of the marginal probability $\theta_k$ in equation (\ref{eq:prob_theta}) is needed.
We consider here the following H\'{a}jek-type estimator (cfr., among the others, \cite{hernrob06}), that is a slight modification of the commonly used inverse probability weighted estimator though with improved statistical properties:
\begin{eqnarray}
\widehat{\theta}^H_k =\frac{1}{ \sum_{i=1}^{n}  I_{(T_{i} =k)} \widehat{p}_k ( \bm{x}_i )^{-1}}  \sum_{i=1}^{n} I_{(Y_{i} = 1)} I_{(T_{i} =k)} \widehat{p}_k ( \bm{x}_i )^{-1} ; \;\;
	k=0, \, 1.
\label{eq:estim_theta-hajek}
\end{eqnarray}
As a natural competitor of (\ref{eq:estim_theta-hajek}),
we also consider the inverse probability weighted estimator ({\em i.e.} the Horvitz-Thompson-type estimator) as in \cite{luncdav04}:
\begin{eqnarray}
	\widehat{\theta}^{HT}_k = \frac{1}{n} \sum_{i=1}^{n} I_{(Y_{i} = 1)} I_{(T_{i} =k)} \widehat{p}_k ( \bm{x}_i )^{-1} ; \;\;
k=0, \, 1 .
\label{eq:estim_theta-ht}
\end{eqnarray}

Our first result is that both (\ref{eq:estim_theta-hajek}) and (\ref{eq:estim_theta-ht})
are  consistent estimators of $\theta_k$.
\begin{prop}
\label{prop_consist}
Under assumptions A1-A3, as $n \rightarrow \infty$:
\begin{eqnarray}
& \, & \left \vert \widehat{\theta}_k^H - \theta_k  \right \vert \stackrel{p}{\rightarrow} 0
; \;\;  k=0, \, 1 ;  \label{eq:cons_haj} \\
& \, & \left \vert \widehat{\theta}_k^{HT} - \theta_k  \right \vert \stackrel{p}{\rightarrow} 0
; \;\;  k=0, \, 1 .  \label{eq:cons_ht}
\end{eqnarray}
\end{prop}

The main result of the present section concerns the asymptotic distribution of the estimators (\ref{eq:estim_theta-hajek}),
(\ref{eq:estim_theta-ht}), once properly normalized. Define first the vectors
\begin{eqnarray}
\widehat{\bm{\theta}}^{H} = \begin{bmatrix} \widehat{\theta}^{H}_0  \\ \widehat{\theta}^{H}_1  \end{bmatrix} ,  \;\;
\widehat{\bm{\theta}}^{HT}  = \begin{bmatrix} \widehat{\theta}^{HT}_0  \\ \widehat{\theta}^{HT}_1  \end{bmatrix} ,  \;\;
\bm{\theta}  = \begin{bmatrix}\theta_0  \\ \theta_1 \end{bmatrix} ,
\nonumber
\end{eqnarray}
\noindent and consider the  bivariate r.v.s
\begin{eqnarray}
 \sqrt{n} ( \widehat{\bm{\theta}}^{H}  - \bm{\theta} )  ,   \;\;
 \sqrt{n} ( \widehat{\bm{\theta}}^{HT}  - \bm{\theta}  )  .
\label{eq:multiv_rv-ht}
\end{eqnarray}

\begin{prop}
\label{prop_limit_distr}
Suppose assumptions A1-A3 hold. Then, there exists a sequence of {\em i.i.d.} bivariate random vectors $ \bm{h}^* ( Y_i , \, T_i , \, \bm{X}_i  )$,
with
\begin{eqnarray}
\bm{h}^* ( Y_i , \, T_i , \, \bm{X}_i  ) =    \left [
    \begin{array}{c}
      h^{*}_0 ( Y_i , \, T_i , \, \bm{X}_i )  \\
      h^{*}_1 ( Y_i , \, T_i , \, \bm{X}_i )
      \end{array}
    \right] \nonumber
\end{eqnarray}
\noindent and $E[ h^{*}_k ( Y_i , \, T_i , \, \bm{X}_i ) =0$ having the following properties as $n \rightarrow \infty$:
\begin{itemize}
\item[$1.$] $\sqrt{n} ( \widehat{\bm{\theta}}^{HT} - \bm{\theta} )$ and $n^{-1/2} \sum_{i=1}^{n} \bm{h}^* ( Y_i , \, T_i , \, \bm{X}_i  ) $ possess the same limiting distribution;
\item[$2.$] $\sqrt{n} ( \widehat{\bm{\theta}}^H - \bm{\theta} )$ and $n^{-1/2} \sum_{i=1}^{n} \bm{h}^* ( Y_i , \, T_i , \, \bm{X}_i  ) $ possess the same limiting distribution;
\item[$3.$] If $V ( h^{*}_k ( Y_i , \, T_i , \, \bm{X}_i ) ) >0$, $k=0, \, 1$, then
$n^{-1/2} \sum_{i=1}^{n} \bm{h}^* ( Y_i , \, T_i , \, \bm{X}_i  ) \stackrel{d}{\rightarrow} \bm{W} $
\end{itemize}
\noindent as $n$ goes to infinity, where $\bm{W}$ possesses a bivariate Normal $ N_{2}( \bm{0} , \, \bm{\Sigma} )$ with null mean vector and covariance matrix
\begin{eqnarray}
\bm{\Sigma} =     \left [
    \begin{array}{c c}
      \sigma_{00} & \sigma_{01}  \\
      \sigma_{10} & \sigma_{11}
      \end{array}
    \right] = E \left [  \bm{h}^* ( Y_i , \, T_i , \, \bm{X}_i  ) \, \bm{h}^* ( Y_i , \, T_i , \, \bm{X}_i  )^T  \right ] . \nonumber
\end{eqnarray}
\end{prop}

As a consequence of Proposition \ref{prop_limit_distr}, the estimators $\widehat{\bm{\theta}}^{H} $ and $ \widehat{\bm{\theta}}^{HT} $ possess the
same limiting distribution, and hence they are asymptotically equivalent. For this reason, we will mainly refer to $ \widehat{\bm{\theta}}^{H} $.

Although the main interest of the present paper is in testing for $\Delta = \theta_1 - \theta_0 =0$, we remark that, again as a consequence of the assumptions made, and using the same arguments as in \cite{kim19}, p. 9, the estimators  $\widehat{\theta}^{H}_1 - \widehat{\theta}^{H}_0$,
$\widehat{\theta}^{HT}_1 - \widehat{\theta}^{HT}_0$ are asymptotically efficient, in the sense that they attain the efficiency bound in
\cite{hahn98}, \cite{kim19}.

\subsection{Testing for treatment effect}
\label{sec:test_treatm_effect}

The primary goal of the present section is to construct a test for the hypothesis problem
(\ref{eq:hp_absence_treatm}), {\em i.e.} a test for the absence of treatment effect.
Define $\Delta = \theta_1 -  \theta_0 $. As already remarked, testing for the absence of treatment effect reduces to the following hypothesis problem
\begin{eqnarray}
\left \{
\begin{matrix}
& H_0 : &   \Delta =0 \\
& H_1 : &   \Delta \neq 0
\end{matrix}
\right .
\label{eq:hp_absence_treatm-b}
\end{eqnarray}
A ``natural'' test-statistic for the above hypotheses problem is
\begin{eqnarray}
D_n = \widehat{\theta}_1^H - \widehat{\theta}_0^H  = \bm{a}^T \widehat{\bm{\theta}}^H  \nonumber
\end{eqnarray}
\noindent where $\bm{a}$ is the vector of components $1$ and $-1$. The limiting distribution of $D_n$ is easily obtained from
Proposition \ref{prop_asympt_norm} which is an immediate consequence of the continuous mapping Theorem.

\begin{prop}
\label{prop_asympt_norm}
Suppose assumptions A1-A3 hold. The following two statements hold.
\begin{itemize}
\item[$1.$] $\sqrt{n} ( D_n - \Delta ) \stackrel{d}{\rightarrow} N(0, \, \sigma^2 )$ as $n \rightarrow \infty$, with
$\sigma^2 = \bm{a}^T \bm{\Sigma} \bm{a}$.
\item[$2.$]
Under $H_0$,  $\sqrt{n} D_n \stackrel{d}{\rightarrow} N(0, \, \sigma^2_0 )$ as $n \rightarrow \infty$.
\end{itemize}
\end{prop}

The asymptotic variance $\sigma^2$ is unknown, and must be estimated on the data. A simple technique is jackknife, based on the systematic omission of each observation from the sample data, on calculating the corresponding estimates of $\Delta$,
 $D_{n-1,(-i)}$, $i = 1,...,n$,
and on computing their average. From the approximation of $D_n$
by a $U$-statistic developed in the proof of Proposition \ref{prop_asympt_norm} (see Appendix), it is not difficult to see that
jackknife estimator is asymptotically equivalent to the variance estimator of $U$-statistics proposed by \cite{sen60} and hence it is consistent. On the other hand, since for each
sub-sample of size $n-1$ the construction of a BN is required, this technique may be
computationally cumbersome, especially for large $n$.
A different approach may be developed by exploiting  (in a different context) ideas in
\cite{hirano03}.
Define $\theta_k(\bm{x}) = P(Y_i = 1|T_i = k, \bm{X}_i = \bm{x})$, $k=0,1$, and
$\widehat{\theta}_k(\bm{x})$ its corresponding estimator obtained with the same approach used for estimating the propensity score (in our case by using a BN). 

Moreover, define
\begin{eqnarray}
\widehat{h}_{i1}& = &\left (\frac{I_{(T_i=1)}}{\widehat{p}(\bm{x}_i)} I_{(Y_{i} =1)}  -  \widehat{\theta}_1 \right) - \frac{\widehat{\theta}_1(\bm{x_i})}{\widehat{p}(\bm{x}_i)} \left(I_{(T_i=1)} - \widehat{p}(\bm{x}_i)\right )  \nonumber \\
\widehat{h}_{i0}& = &\left (\frac{I_{(T_i=0)}}{1-\widehat{p}(\bm{x}_i)} I_{(Y_{i} =1)}  -  \widehat{\theta}_0 \right) - \frac{\widehat{\theta}_0(\bm{x_i})}{1-\widehat{p}(\bm{x}_i)} \left(I_{(T_i=0)} - (1-\widehat{p}(\bm{x}_i))\right )
 \nonumber
\end{eqnarray}
and
\begin{eqnarray}
\widehat{d}_{i}=\widehat{h}_{i1}-\widehat{h}_{i0}, i=1,...,n.
 \nonumber
\end{eqnarray}
Using the same approach as in \cite{hirano03}, it is possible to see that
\begin{eqnarray}
\widehat{\sigma}_{n}^2=\frac{1}{n}\sum_{i=1}^n(\widehat{d}_{i})^2
 \nonumber
\end{eqnarray}
is a consistent estimator of $\sigma^2$.
As a consequence, in order to test for the presence of treatment effect, a simple procedure consists in constructing the following confidence interval at level $1-\alpha$ for $\Delta$
\begin{eqnarray}
	\left [ D_{n} -z_{\alpha /2} \frac{\widehat{\sigma}_n}{\sqrt{n}} , \;
	 D_{n} +z_{\alpha /2} \frac{\widehat{\sigma}_n}{\sqrt{n}} \right ] \label{eq:conf-int-Delta},
\end{eqnarray}
where $\widehat{z}_{p}$ is the $(1-p)$-quantile of the Standard Normal distribution, and in rejecting $H_0$ whenever the interval (\ref{eq:conf-int-Delta} ) does not contain $0$.

Finally, as an approximated $p$-value for the above testing procedure, we may take
$2 ( 1- \Phi ( \sqrt{n} D_n / \widehat{\sigma}_n ) )$, where $\Phi$ is the Standard Normal distribution function.

\section{Estimation of treatment effects under misspecification of the propensity score model}
\label{sec:misspecification}
As seen in the previous sections, Bayesian Networks are an excellent tool for estimating propensity scores, in view of their ``universal'' consistency in eqn. (\ref{eq:cons_pk_BN}). Asymptotically, BNs would ensure the removal of all bias in the potential outcomes probabilities estimates, since they are both universally consistent and parsimonious in terms of parameters used. However, under a limited sample size, or when structural learning is not used, or in case of omission of relevant covariates, there could be an uncorrect specification of propensity scores.



The goal of the present section is to study the behaviour of estimators $\widehat{\theta}^{HT}_k$, $\widehat{\theta}^{H}_k$ when the model
for propensity scores  is uncorrectly specified. More specifically, assume that the statistical model
\begin{eqnarray}
p_k (  \bm{x} ; \, \bm{\beta} ) = P( T=k \vert   , \bm{X} = \bm{x} ; \, \bm{\beta} ) , \; k=0, \, 1 ;  \;\; \bm{\beta} \in \bm{\Upsilon}
\label{eq:param-misspecif}
\end{eqnarray}
\noindent is adopted for propensity scores, where $ \bm{\beta} $ is a multidimensional parameter and $\bm{\Upsilon}$ is the corresponding
parameter space. In the sequel, we focus attention on the case the model (\ref{eq:param-misspecif}) is misspecified for the propensity score $p_k(\bf x)$, {\em i.e.} it does not contain, $p_k^{TRUE} ( \bm{x} )$.


Let $\widehat{\bm{\beta}}_n$ be the MLE of $\bm{\beta}$. From \cite{white82}, Th. 2.2, it is not difficult to see that, under mild regularity
conditions on the model (\ref{eq:param-misspecif}) (they are essentially Wald's conditions for consistency of MLEs),
as $n$ increases  $\widehat{\bm{\beta}}_n$ converges a.s. to
\begin{eqnarray}
\bm{\beta}^* = {\mathrm{argmin}} \left \{ E \left [ p_1 ( \bm{x} ) \log \frac{p_1 ( \bm{x} )}{p_1 (  \bm{x} ; \, \bm{\beta} )}
+ (1- p_1 ( \bm{x} ) ) \log \frac{1- p_1 ( \bm {x} )}{1- p_1 (  \bm{x} ; \, \bm{\beta} )}
\right ] \right \}. \nonumber
\end{eqnarray}
\noindent In other words, $p_1 (  \bm{x} ; \, \bm{\beta}^* )$ minimizes the (average) Kullback-Leibler divergence of the adopted
probability distribution for the propensity score from the true one.

Let us now consider the
H{\'a}jek and Horvitz-Thompson estimators of $\theta_k$ when the propensity score model is uncorrectly specified, namely
\begin{eqnarray}
\widehat{\theta}^H_k & = & \frac{1}{ \sum_{i=1}^{n}  I_{(T_{i} =k)} p_k ( \bm{x}_i ; \, \widehat{\bm{\beta}}_n )^{-1}}  \sum_{i=1}^{n} I_{(Y_{i} = 1)} I_{(T_{i} =k)} p_k ( \bm{x}_i ; \, \widehat{\bm{\beta}}_n )^{-1} \nonumber \\
\widehat{\theta}^{HT}_k & = & \frac{1}{n} \sum_{i=1}^{n} I_{(Y_{i} = 1)} I_{(T_{i} =k)} p_k ( \bm{x}_i ; \, \widehat{\bm{\beta}}_n )^{-1} \nonumber
\end{eqnarray}
Their limiting behaviour is studied in Proposition \ref{HT_H_asympt_misspec}.

\begin{prop}
\label{HT_H_asympt_misspec}
Suppose that $p_k ( \bm{x} ; \, \bm{\beta} )$ is a continuous function of $\bm{\beta}$ for each fixed $\bm{x}$, that assumptions $A1$-$A3$ in \cite{white82} and $A1$, $A2$ are satisfied, and that there exists $\delta >0$ such that $\delta \leq
p_k ( \bm{x} ; \, \bm{\beta}^* ) \leq 1- \delta$ for each $\bm{x}$.
Then, the following two statements hold
\begin{small}
\begin{eqnarray}
\widehat{\theta}^{HT}_k - \theta_k & \stackrel{p}{\rightarrow} & E \left [ \theta_k ( \bm{X} ) \left (
\frac{p_{k} ( \bm{X} )}{p_k (  \bm{X} ; \, \bm{\beta}^* )} -  1 \right )
\right ] \;\; {\mathrm{as}} \; n \rightarrow \infty ; \label{eq_ht_misspec} \\
\widehat{\theta}^{H}_k - \theta_k & \stackrel{p}{\rightarrow} &
\frac{1}{E \left [ \frac{p_{k} ( \bm{X} )}{p_k (  \bm{X} ; \, \bm{\beta}^* )} \right ]}
E \left [ ( \theta_k ( \bm{X} ) - \theta_k ) \left (
\frac{p_{k} ( \bm{X} )}{p_k (  \bm{X} ; \, \bm{\beta}^* )} -  1 \right ) \right ]
\;\; {\mathrm{as}} \; n \rightarrow \infty \nonumber \\ \label{eq_h_misspec}
\end{eqnarray}
\end{small}
\noindent for $k=0$, $1$.
\end{prop}

The comparison of (\ref{eq_ht_misspec}) and (\ref{eq_h_misspec}) makes it evident that $\widehat{\theta}^{HT}_k $ and
$\widehat{\theta}^{H}_k $ are asymptotically equivalent only when the model for propensity score is correctly specified.
In case of misspecification, their asymptotic behaviour is different.
From (\ref{eq_ht_misspec}) it appears that the limit in probability of $\widehat{\theta}^{HT}_k - \theta_k$ is small {\em only}
when $p_k (  \bm{x} ; \, \bm{\beta}^* )$ is close to $p_{k} ( \bm{x} )$, {\em i.e.} when misspecification is negligible.
On the other hand, (\ref{eq_h_misspec}) shows that  the limit in probability of $\widehat{\theta}^{H}_k - \theta_k$ is small
if {\em either} $p_k (  \bm{x} ; \, \bm{\beta}^* )$ is close to $p_{k} ( \bm{x} )$ {\em or} $\vert \theta_k ( \bm{x} ) - \theta_k \vert $
is small, namely when  $\theta_k (1 \vert \bm{x} )$ does not vary too much around $\theta_k $.
In addition, from (\ref{eq_h_misspec}) and taking into account that $E [ \theta_k ( \bm{X} ) ] = \theta_k$, the inequality
\begin{eqnarray}
\left \vert \frac{1}{E \left [ \frac{p_{k} ( \bm{X} )}{p_k (  \bm{X} ; \, \bm{\beta}^* )} \right ]}
E \left [ ( \theta_k ( \bm{X} ) - \theta_k ) \left (
\frac{p_{k} ( \bm{X} )}{p_k (  \bm{X} ; \, \bm{\beta}^* )} -  1 \right ) \right ] \right \vert
\leq \sup_{\bm{x}} \vert \theta_k ( \bm{x} ) - \theta_k \vert \nonumber
\end{eqnarray}
\noindent is obtained. Hence, the limit in probability of $ \vert \widehat{\theta}^{H}_k - \theta_k \vert$ is bounded. The same does not happen
for the Horvitz-Thompson estimator $\widehat{\theta}^{HT}_k$, because the expectation
\begin{eqnarray}
E \left [ \theta_k ( \bm{X} ) \left (
\frac{p_{k} ( \bm{X} )}{p_k (  \bm{X} ; \, \bm{\beta}^* )} -  1 \right )
\right ] \nonumber
\end{eqnarray}
\noindent is unbounded. This means that
$\widehat{\theta}^{H}_k$ is preferable for being less prone to misspecification of propensity scores model
if compared to $\widehat{\theta}^{HT}_k$.

\section{Application}
\label{sec:mot-ex}
In this section, we will apply the proposed approach discussed in the previous sections to a clinical application. We have considered data from 6478 prostate cancer patients who underwent radical prostatectomy at San Raffaele Hospital (Milan, Italy). The goal is to evaluate whether receiving neoadjuvant hormonal therapy (NEOadjHT) before radical prostatectomy has an effect on the decision to perform lymphadenectomy during the surgery. Since  data are observational, the decision to administer neoadjuvant hormonal therapy is based solely on patient characteristics. However, since these variables also affect the outcome, there may be bias in the estimates of treatment effects on the outcome. The variables that could influence the decision on neoadjuvant hormonal therapy are age, body mass index (BMI), Charlson Comorbidity Index (CCI), biopsy Gleason score (bxgg), clinical stage (clinstage), and total PSA (tpsa). These variables also affect the decision to perform lymphadenectomy. Since categorical or categorized covariates are commonly used for decision-making in clinical practice, we have categorized all the quantitative covariates using clinical cut-offs in this study.

\begin{figure}[!h]
\begin{center}
\includegraphics[width=0.46\textwidth]{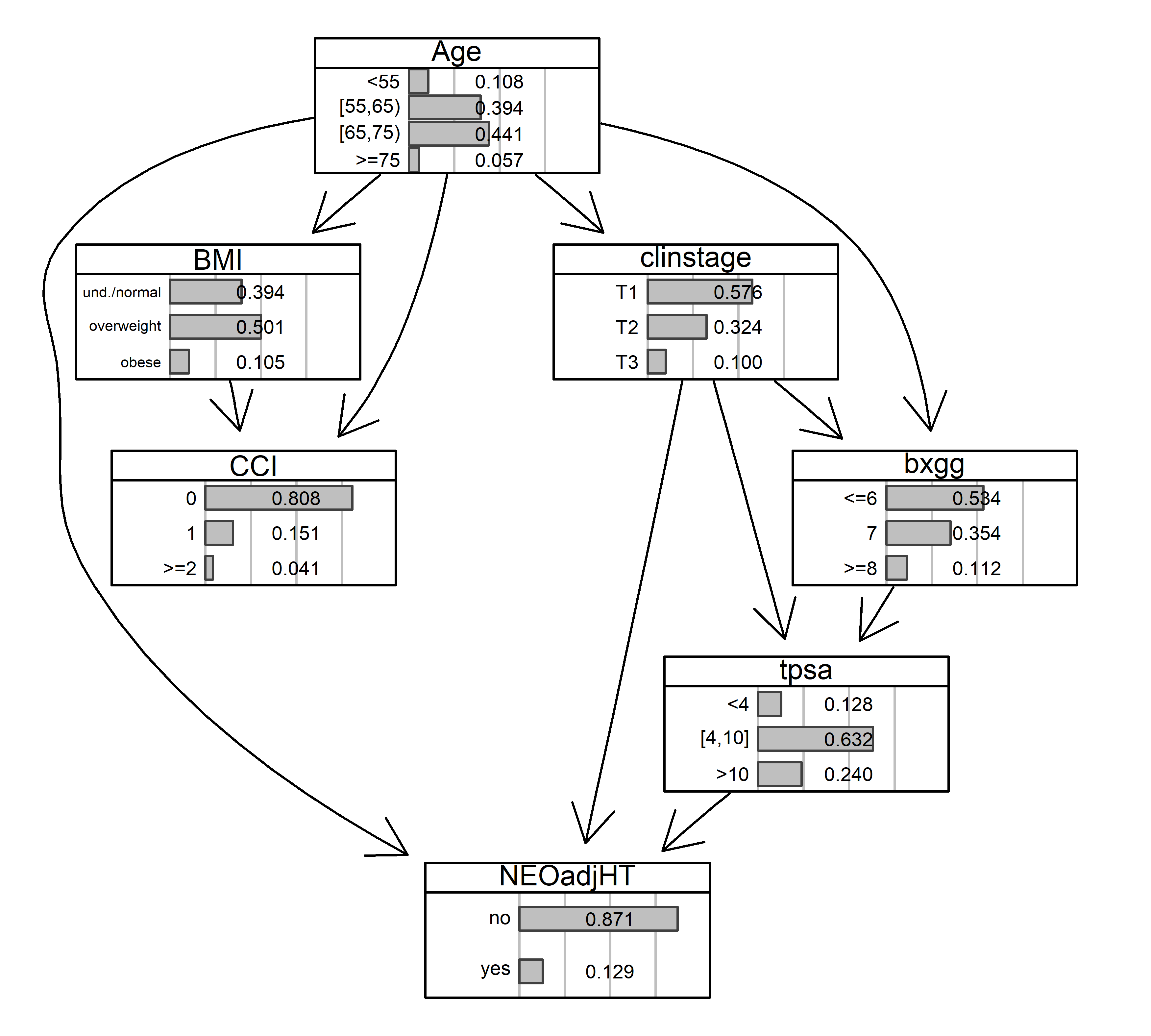}
\includegraphics[width=0.46\textwidth]{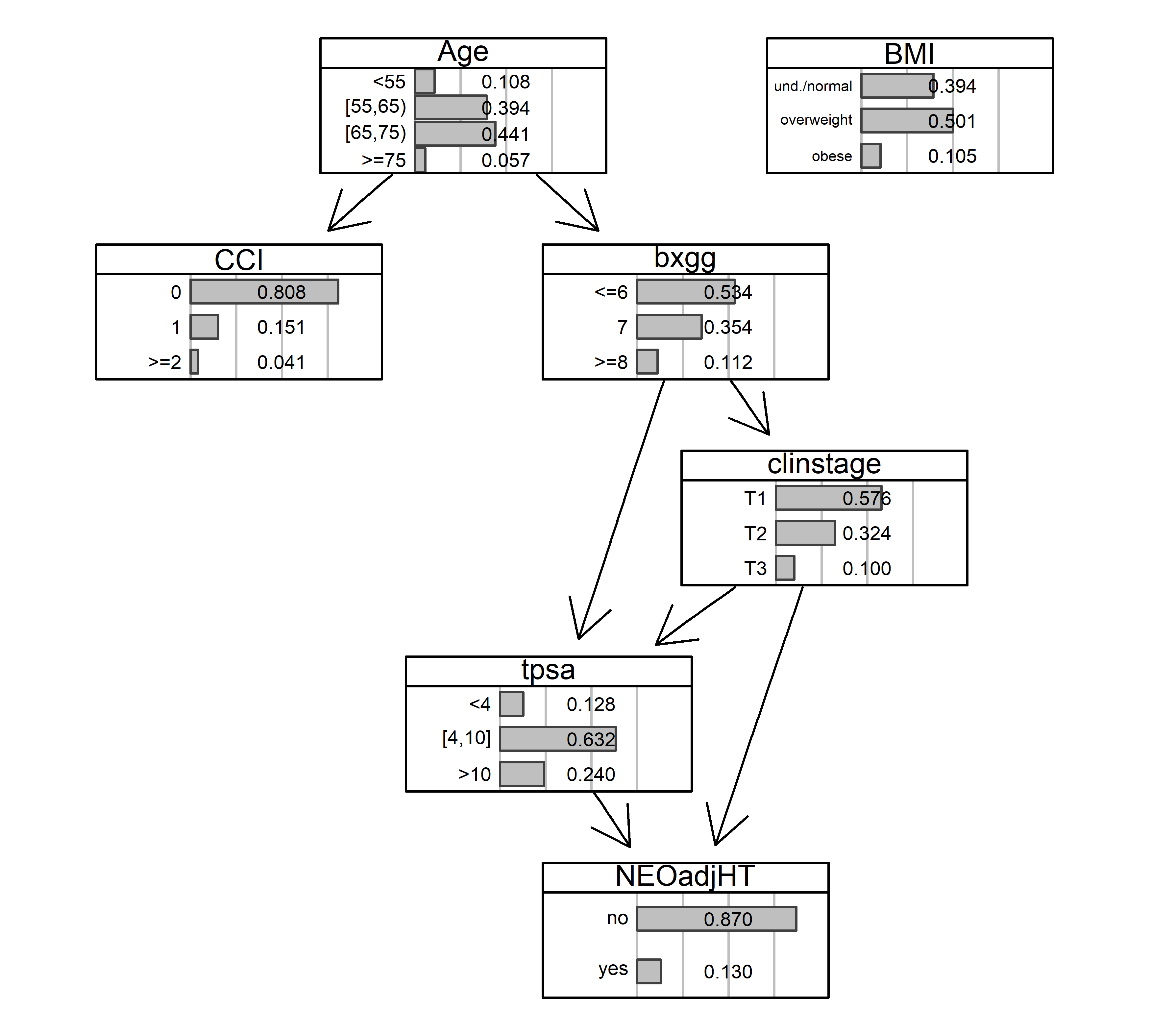}
\end{center}
\caption{Bayesian networks obtained on the real prostate cancer data through the Tabu  greedy search (TABU) algorithm with the AIC (on the left hand side) and BIC  (on the right hand side) score functions. The variables in the data are: Body Mass Index (BMI),  Charlson Comorbidity Index (CCI), biopsy Gleason score (bxgg), clinical stage (clinstage) and total PSA (tpsa).}
\label{fig:1}       
\end{figure}

The Bayesian networks (BNs) displayed in Figure \ref{fig:1} were estimated on real prostate cancer data using the Tabu greedy search (TABU) algorithm with AIC and BIC score functions, respectively. It can be observed that, in both cases, the estimated BN supports the presence of interaction terms among some covariates for the estimation of the propensity score. Therefore, in this real setting, the use of BN should provide an advantage for a good estimation of the propensity score and subsequently for the ATE estimation.

\begin{figure}[!h]
\begin{center}
\includegraphics[width=0.9\textwidth]{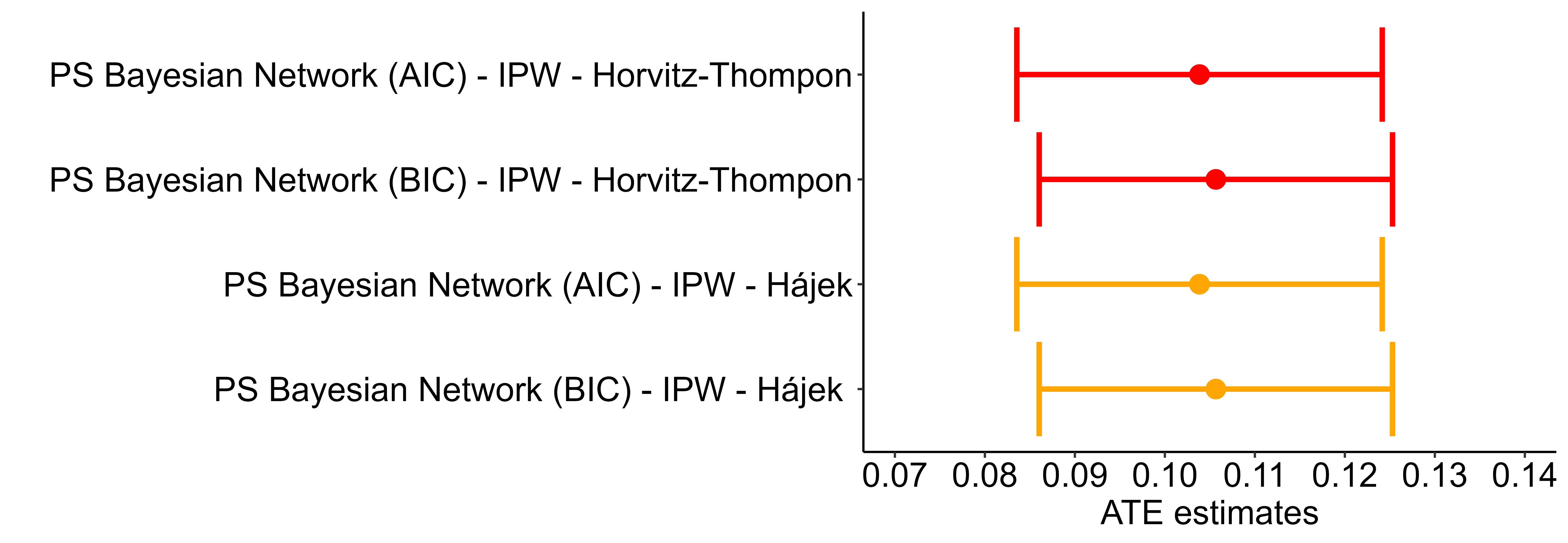}
\end{center}
\caption{ATE estimates and 95$\%$  confidence intervals (bars). }
\label{fig:2real}       
\end{figure}

Figure \ref{fig:2real} presents the ATE estimates, along with 95\% confidence intervals, using the H{\'a}jek-type (\ref{eq:estim_theta-hajek}) and the Horvitz-Thompson-type (\ref{eq:estim_theta-ht}) estimators. The propensity score was estimated by a Bayesian network based on either AIC or BIC score functions. All methods gave confidence intervals that did not include zero. Hence, there is sufficient evidence to reject the hypothesis that neoadjuvant hormonal therapy before radical prostatectomy did not have an effect on the decision to perform lymphadenectomy during the surgery. In this application, the differences in the ATE estimate among the different methodologies were negligible.
To evaluate the goodness of the proposed approach and to delve deeper into the impact on the estimation of ATE of different PS estimations combined with the usage of the H{\'a}jek-type (\ref{eq:estim_theta-hajek}) and the Horvitz-Thompson-type (\ref{eq:estim_theta-ht}) estimators,
a simulation study has been designed in Section \ref{sec:empir-RWD}. It mimics the characteristics of this real dataset while varying the sample size.

\section{Empirical evidence from data mimicking prostate cancer real data}
\label{sec:empir-RWD}
In this section,
we use artificial data that mimic the characteristics of a real dataset of prostate cancer patients, described in the previous section. Our main goal is to provide empirical evidence with finite sample sizes $n$ of the theoretical asymptotic results in Sections 3 to 4.

\subsection{Simulation study plan}
\label{subs:RWD-sim-plan}
The simulated data were generated in order to mimic the prostate cancer real data of Section \ref{sec:mot-ex}. Treatment assignment and covariates have been generated based on the estimated BN by using the Tabu  greedy search (TABU) algorithm with AIC score function (Figure \ref{fig:1} left).

The binary potential outcomes $Y_{(0)}$ and $Y_{(1)}$ were then simulated through logistic models including the whole set of covariates.
Again, values of  the model coefficients were equal to the observed coefficients estimated on the prostate cancer data. In detail, if $B$ denotes the Bernoulli distribution,
\begin{eqnarray}
Y_{(k)}|X_c \sim B(P(Y_{(k)}=1|  \bm{X}_c)) \;\;\;\mbox{for }\;k=0,1,
 \nonumber
\end{eqnarray}
where $Y_{(k)}=1$ if lymphadenectomy is performed and $Y_{(k)}=0$ otherwise, for $k=0,1$, $\bm{X}_c$ denotes the vector of covariates, and
\begin{eqnarray}
logit\left( P(Y_{(0)}=1|\bm{X}_c)\right) = \alpha_0 +\bm{\beta}^T \bm{X} _c,   \nonumber \\
logit\left( P(Y_{(1)}=1|\bm{X}_c) \right) = \alpha_0 +\alpha_1+ \bm{\beta}^T \bm{X}_c.
 \nonumber
\end{eqnarray}
In order to compute the true ATE for the evaluation of the methods, the probabilities $\theta_k=P(Y_{(k)}=1)$, $k=0,1$, were obtained by marginalizing $P(Y_{(k)}=1 | \bm{X}_c)$ over $\bm{X}_c$.
Fixing the parameters $\alpha_0$, $\alpha_1$ and $\bm{\beta}^T$ based on the real data, the corresponding ATE is equal to $0.094$.  The outcome $Y$ was generated according to eqn. (\ref{eq:observed_outc}),
with treatment $T$ referring to the neoadjuvant hormonal therapy:
$T=1$ if the therapy was done, and $T=0$ otherwise.
In this set up mimicking real data, we considered four sample sizes ($n$ = 500, 1000, 2500, 5000) and 1000 Monte Carlo (MC) runs for each sample size.

To estimate the propensity score we used 
Bayesian Networks learned with the Tabu greedy search  algorithm with either AIC score function (BN AIC) or BIC score function (BN BIC). The estimated propensity score was then used to obtain the estimate of ATE and the related 95\% confidence interval (CI) through two inverse probability weighting  estimators for $\theta_k$: the Horvitz-Thompson estimator $\widehat{\theta}^{HT}_k$ in (\ref{eq:estim_theta-ht}) and the H{\'a}jek estimator $\widehat{\theta}^H_k$ in (\ref{eq:estim_theta-hajek}).

\subsection{Results and discussion}
\label{subs:RWD-results}
\begin{figure}[H]
\begin{center}
\includegraphics[width=0.9\textwidth]{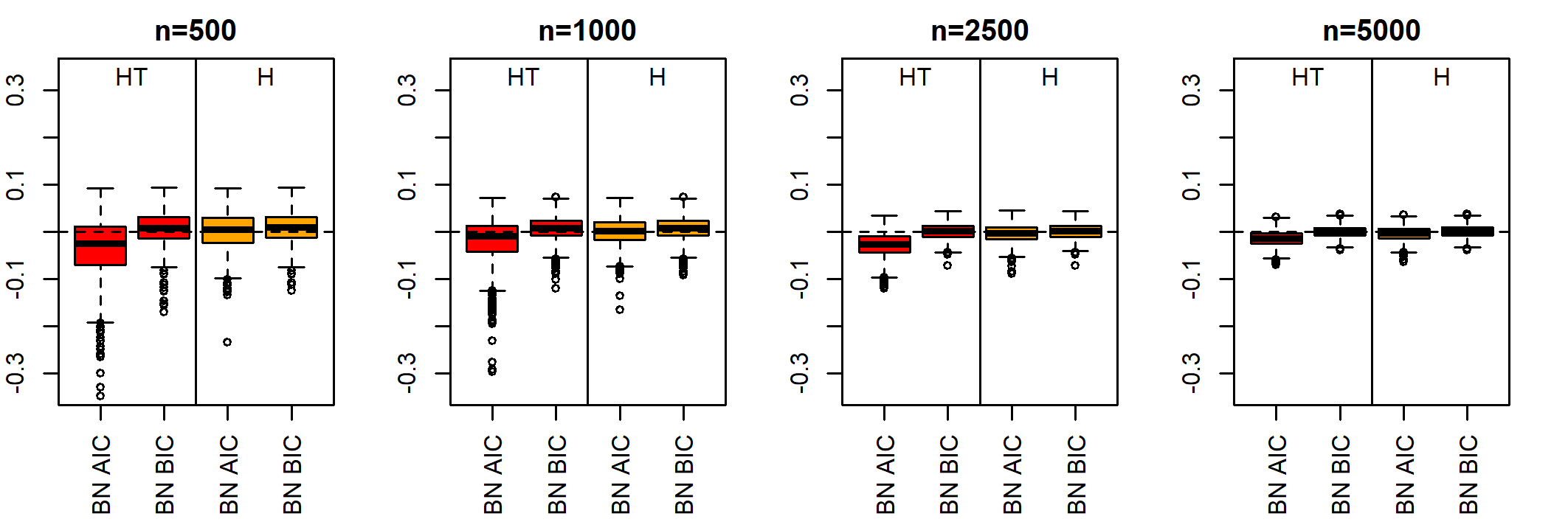}
\end{center}
\caption{Bias of the  ATE estimators obtained with the two 
BNs for PS estimation on the simulated data mimicking the real prostate cancer data. The bias is defined as the difference between the estimated and the true ATE. HT = Horvitz-Thompson estimator, H = H{\'a}jek estimator.}
\label{fig:SimRealiBias}       
\end{figure}

In Figure \ref{fig:SimRealiBias}, we present box-plots that summarize the differences in bias among the ATE estimators. The closer the median
bias over simulations to zero, the higher the estimation accuracy. The lower the variability of the distribution, the higher the estimator efficiency.
As far as the Horvitz-Thompson estimator is concerned, the ATE estimator's bias using the propensity score estimated by BN BIC is generally closer to zero than the bias obtained through BN AIC.
In this latter case we have less precision and a tendency to underestimate the ATE.
In case of H{\'a}jek estimator, both BN AIC and BN BIC exhibit a good performance. Increasing sample sizes uniformly improve estimation.

Table~\ref{tab:resultsReal} shows two additional evaluation metrics for increasing sample sizes: 1) the empirical coverage (EC) of 95\% CIs, computed as the proportion of CIs in eqn. (\ref{eq:conf-int-Delta}) that include the true ATE (0.094) within 1000 runs and 2) the empirical rejection rate (ERR) of the test for no treatment effect specified in eqn. (\ref{eq:hp_absence_treatm-b}) at the $\alpha=0.05$ level, computed as the proportion of CIs in eqn. (\ref{eq:conf-int-Delta}) not including zero.

In case of Horvitz-Thompson estimator, BN BIC gives better results than BN AIC in terms of both emprical coverage and its closeness to the nominal level. With the H{\'a}jek estimator of ATE, the largest coverage is obtained by using the BN AIC propensity score, with a tendency to provide conservative CIs (i.e., coverage larger than 95\%). The coverage obtained by using the BN BIC approach is the closest to the nominal 95\%.

With both types of ATE estimators, the larger proportion of rejection of the null hypothesis is achieved using the BN BIC rather than BN AIC propensity score. Moreover, this proportion increases to 1 as the sample size increases. 


\begin{table}[!bth]
\caption{Empirical coverage (EC) of CIs in eqn. (\ref{eq:conf-int-Delta}) and  empirical rejection rate (ERR) of no treatment effect  test (\ref{eq:hp_absence_treatm-b}),  for increasing sample size $n$,  for the two PS estimation approaches (BN AIC and BN BIC) and the two ATE estimators, by using
the simulated data mimicking the real prostate cancer data.}
\label{tab:resultsReal}       
\small
\begin{tabular}{llcccc}
\hline\noalign{\smallskip}
&& \multicolumn{2}{c}{Horvitz-Thompson}&\multicolumn{2}{c}{H{\'a}jek}\\		
	n& Methods	& EC&	 ERR	 & EC&	 ERR\\
\hline
500	&	BN AIC	&	0.862	&	0.364	&	0.999	&	0.574	\\
500	&	BN BIC	&	0.976	&	0.73	&	0.985	&	0.736	\\
\hline
1000	&	BN AIC	&	0.852	&	0.699	&	0.985	&	0.841	\\
1000	&	BN BIC	&	0.972	&	0.932	&	0.975	&	0.933	\\
\hline
2500	&	BN AIC	&	0.779	&	0.813	&	0.984	&	0.955	\\
2500	&	BN BIC	&	0.973	&	0.999	&	0.975	&	0.999	\\
\hline
5000	&	BN AIC	&	0.868	&	0.986	&	0.963	&	0.994	\\
5000	&	BN BIC	&	0.946	&	1	&	0.945	&	1	\\
\hline
\end{tabular}
\end{table}

\section{Conclusions}
\label{sec:conclusion}
In this paper, a new method for estimating and testing the Average Treatment Effect (ATE) for binary outcomes using Bayesian Networks (BNs) for Propensity Score (PS) estimation is proposed. We consider both Horvitz-Thompson (HT) and H{\'a}jek (H) type estimators of ATE, with the second being a modification of the first known to be more stable for small to moderate sample sizes. Asymptotic properties of the ATE estimators are derived by the statistical properties of BNs.
The main conclusions of the present paper can be summarized as follows:

\begin{itemize}
\item[1.] Estimating PS by BNs is particularly important when the dependence structure for treatments outcome and covariates is complex.
BN structure can be learned directly from data (eg. via structural learning) involving treatment assignment and covariates without necessarily imposing a priori functional relationships. Under the assumption of discrete covariates, BNs allow us to produce maximum likelihood (ML) estimators, which are asymptotically unbiased and efficient as well as normally distributed. 
\item[2.] As a consequence of point 1, the use of BNs ensures the efficiency of both Horvitz-Thompson and H{\'a}jek estimators of ATE. It also allows us to define a test for the absence of treatment effect.
\item[3.] 
As theoretically explained in Section~\ref{sec:misspecification} and highlighted from our simulation study, the H{\'a}jek estimator seems recommendable over the more familiar Horvitz-Thompson estimator for inverse probability weighting to estimate the ATE, for producing interval estimates and for testing null treatment effect hypothesis.
\item[4.] As the sample size $n$ increases, BNs show a clear improvement, according to its ``universal'' consistency property given in eqn. (\ref{eq:cons_pk_BN}) and discussed in Section \ref{sec:misspecification}

\end{itemize}

\backmatter

\begin{appendices}

\section*{Appendix}

\begin{Lemma}
\label{lemma_1}
Under Assumptions A1, A3, the following results hold.
\begin{eqnarray}
\sup_{\mathbf{x}} \left \vert  \widehat{p}_k ( \bm{x} ) - p_k ( \bm{x} ) \right \vert & \stackrel{p}{\rightarrow} & 0 \;\;
{\mathrm{as}} \; n \rightarrow \infty \;\; \forall \, k=0, \, 1; \nonumber \\
\sup_{\bm{x}} \left \vert  \frac{1}{\widehat{p}_k ( \bm{x} )} - \frac{1}{p_k ( \bm{x} )}
 \right \vert & \stackrel{p}{\rightarrow} & 0 \;\;
{\mathrm{as}} \; n \rightarrow \infty \;\; \forall \, k=0, \, 1  . \nonumber
\end{eqnarray}
\end{Lemma}
\begin{proof} Immediate consequence of (\ref{eq:cons_pk_BN}) and the finiteness of possible values of $\bm{x}$.
\end{proof}

\begin{Lemma}
\label{lemma_2}
Under Assumptions A1-A3:
\begin{eqnarray}
\frac{1}{n} \sum_{i=1}^{n} \frac{I_{(T_{i} =k)}}{\widehat{p}_k ( \bm{X}_{i} )} \stackrel{p}{\rightarrow} 1
\;\;
{\mathrm{as}} \; n \rightarrow \infty  ; \;\;  k=0, \, 1  . \label{eq:wlln_prop}
\end{eqnarray}
\end{Lemma}
\begin{proof}
First of all, we may write
\begin{eqnarray}
\frac{1}{n} \sum_{i=1}^{n} \frac{I_{(T_{i} =k)}}{\widehat{p}_k ( \bm{X}_{i} )} =
\frac{1}{n} \sum_{i=1}^{n} \left ( \frac{1}{\widehat{p}_k ( \bm{X}_{i} )} -
\frac{1}{p_k ( \bm{X}_{i} )} \right )
 I_{(T_{i} =k)} +
\frac{1}{n} \sum_{i=1}^{n} \frac{I_{(T_{i} =k)}}{p_k ( \bm{X}_{i} )} .
\label{eq:dcp1}
\end{eqnarray}
In addition, since
\begin{eqnarray}
E \left [ \frac{I_{(T =k)}}{p_k ( \bm{X} )}  \right ] =
E \left [ \frac{1}{p_k ( \bm{x} )}
E \left [ I_{(T =k)} \left \vert \right .  \bm{X} = \bm{x} \right ] \right ] = 1
\nonumber
\end{eqnarray}
\noindent from the Weak Law of Large Numbers (WLLN) it is seen that
\begin{eqnarray}
\frac{1}{n} \sum_{i=1}^{n} \frac{I_{(T_{i} =k)}}{p_k ( \bm{X}_{i} )}
 \stackrel{p}{\rightarrow} 1 \;\;
{\mathrm{as}} \; n \rightarrow \infty . \nonumber
\end{eqnarray}
Finally, observing that
\begin{eqnarray}
\left \vert
\frac{1}{n} \sum_{i=1}^{n} \left ( \frac{1}{\widehat{p}_k ( \bm{X}_{i} )} -
\frac{1}{p_k ( \bm{X}_{i} )} \right )
 I_{(T_{i} =k)} \right \vert & \leq &
\sup_{\bm{x}} \left \vert  \frac{1}{\widehat{p}_k ( \bm{x} )} - \frac{1}{p_k ( \bm{x} )}
 \right \vert
\nonumber
\end{eqnarray}
\noindent conclusion (\ref{eq:wlln_prop}) immediately follows from Lemma $\ref{lemma_1}$ and
(\ref{eq:dcp1}).
\end{proof}

\begin{proof}[\textbf{Proof of Proposition \ref{prop_consist}}]
Define first the ``pseudo-estimator'' of $\theta_k $:
\begin{eqnarray}
\widetilde{\theta}_k = \frac{1}{n} \sum_{i=1}^{n} \frac{I_{(T_{i} =k)}}{p_{k} ( \bm{X}_{i} )} I_{(Y_{i} =1)} , \;\; k=0, \, 1.
\nonumber
\end{eqnarray}
From the WLLN, and using A2, it is immediate to see that
\begin{small}
\begin{eqnarray}
\widetilde{\theta}_k  & \stackrel{p}{\rightarrow} & E \left [  \frac{I_{(T_{i} =k)}}{p_{k} ( \bm{X}_{i} )} I_{(Y_{i} = 1)} \right ] =
E \left [ \frac{1}{p_k ( \bm{x} )}
E \left [ I_{(T =k)} \left \vert \right .  \bm{X} = \bm{x} \right ]
E \left [ I_{(Y_{(k)} =1)} \left \vert \right .  \bm{X} = \bm{x} \right ]
 \right ] \nonumber \\
& = & \theta_k , \;\; k=0, \, 1 \label{eq_convp1}
\end{eqnarray}
\end{small}
\noindent as $n \rightarrow \infty$.

The consistency of the Horviz-Thompson estimator $\widehat{\theta}^{HT}_k (1)$ is then an immediate consequence of
\begin{eqnarray}
\left \vert \widehat{\theta}^{HT}_k  - \widetilde{\theta}_k  \right \vert & = &
\left \vert \frac{1}{n} \sum_{i=1}^{n} \left ( \frac{1}{\widehat{p}_k ( \bm{X}_i )} -
\frac{1}{p_k ( \bm{X}_i )}    \right ) I_{(T_{i} =k)} I_{(Y_{i} =1)} \right \vert \nonumber \\
& \leq & \sup_{\bm{x}} \left \vert  \frac{1}{\widehat{p}_k ( \bm{x} )} - \frac{1}{p_k ( \bm{x} )}
 \right \vert \nonumber
\end{eqnarray}
\noindent and Lemma $\ref{lemma_1}$.

As far as the H{\'a}jek-type estimator $\widehat{\theta}_k (1)$ is concerned, it can be proved as a consequence of the relationship
\begin{eqnarray}
\widehat{\theta}_k - \widetilde{\theta}_k  & = &
\frac{1}{\frac{1}{n} \sum_{j=1}^{n}  \frac{I_{(T_{j} =k)}}{\widehat{p}_k ( \bm{X}_j )}} \,
\left \{ \frac{1}{n} \sum_{i=1}^{n} \left ( \frac{1}{\widehat{p}_k ( \bm{X}_j )} -
\frac{1}{p_k ( \bm{X}_j )}    \right ) I_{(T_{i} =k)} I_{(Y_{i} =1)} \right \} \nonumber \\
& + & \left \{ \left ( \frac{1}{n} \sum_{j=1}^{n}  I_{(T_{j} =k)}
\right )^{-1} - 1 \right \} \widetilde{\theta}_k.
\nonumber
\end{eqnarray}
\noindent and of Lemmas $\ref{lemma_1}$, $\ref{lemma_2}$
and (\ref{eq_convp1}).
\end{proof}

\begin{proof}[\textbf{Proof of Proposition \ref{prop_limit_distr}}]
Proof is in principle simple, apart a few complications in notation. Let us start with with the HT-type estimator
(\ref{eq:estim_theta-ht}).
First of all, using standard results for MLEs, it is not difficult to see that
\begin{eqnarray}
\widehat{p}_k ( \bm{x} ) - p_k ( \bm{x} ) = \frac{1}{n} \sum_{i=1}^{n} S_k (  T_i , \, \bm{X}_i ; \, \bm{x} )
+ o_p ( n^{-1} ) \nonumber
\end{eqnarray}
\noindent where  $S_k ( T_i , \, \bm{X}_i ; \, \bm{x} )$ is a linear transformation of the score vector function for $i$th observation. Note that
$E[ S_k ( T_i , \, \bm{X}_i ; \, \bm{x} ) ] =0$. From the above relationship, it is no difficult to see that

\small{
\begin{eqnarray}
\widehat{\theta}^{HT}_k & = &  \frac{1}{n} \sum_{i=1}^{n} I_{(Y_{i} = 1)} I_{(T_{i} =k)} p_k ( \bm{X}_i )^{-1} +
\frac{1}{n} \sum_{i=1}^{n} I_{(Y_{i} = 1)} I_{(T_{i} =k)} \left ( \widehat{p}_k ( \bm{X}_i )^{-1} -  p_k ( \bm{X}_i )^{-1} \right ) \nonumber \\
& = & \frac{1}{n} \sum_{i=1}^{n} I_{(Y_{i} = 1)} I_{(T_{i} =k)}  \left ( p_k ( \bm{X}_i )^{-1} -
p_k ( \bm{X}_i )^{-2} ( \widehat{p}_k ( \bm{X}_i ) - p_k ( \bm{X}_i ) \right )  + O_p ( n^{-1} ) \nonumber \\
& = & \frac{1}{n} \sum_{i=1}^{n} I_{(Y_{i} = 1)} I_{(T_{i} =k)} \left ( p_k ( \bm{X}_i )^{-1} - p_k ( \bm{X}_i )^{-2}
\frac{1}{n} \sum_{j=1}^{n} S_k ( T_j , \, \bm{X}_j ; \, \bm{X}_i ) \right ) + o_p ( n^{-1/2} ) \nonumber \\
& = & \frac{1}{n^{2}} \sum_{i=1}^{n} \sum_{j \neq i} \tau_k ( Y_i , \, T_i , \, \bm{X}_i ;
\, T_j , \,  \bm{X}_j ) + o_p ( n^{-1/2} ) \label{eq:stat-u_1}
\end{eqnarray}}
\normalsize
\noindent where

\begin{eqnarray}
\tau_k ( Y_i , \, T_i , \, \bm{X}_i ;
\, T_j , \,  \bm{X}_j ) = I_{(Y_{i} = 1)} I_{(T_{i} =k)} \left ( p_k ( \bm{X}_i )^{-1} - p_k ( \bm{X}_i )^{-2}
 S_k ( T_j , \, \bm{X}_j , \, \bm{X}_i ) \right ) . \nonumber
\end{eqnarray}

Define next the symmetric kernel
\begin{eqnarray}
h_k ( Y_i , \, T_i , \, \bm{X}_i ; \, Y_j , \, T_j , \, \bm{X}_j ) = \frac{1}{2} \left (
\tau_k ( Y_i , \, T_i , \, \bm{X}_i ; \, T_j , \,  \bm{X}_j ) +
\tau_k ( Y_j , \, T_j , \, \bm{X}_j ; \, T_i , \,  \bm{X}_i )
\right ) \nonumber
\end{eqnarray}
\noindent and consider the $U$-statistic (of degree $2$)
\begin{eqnarray}
U_{n , k}  = \frac{1}{n (n-1)} \sum_{i=1}^{n} \sum_{j \neq i} h_k ( Y_i , \, T_i , \, \bm{X}_i ; \, Y_j , \, T_j , \, \bm{X}_j ) , \;\; k=0, \, 1.
\label{eq:stat-u_2}
\end{eqnarray}
Taking into account that $E[ S_k ( T_i , \, \bm{X}_i , \, \bm{x} ) ] =0$ for every fixed $\bm{x}$ and $k$, it is not difficult
to see that $E[ U_{n , k}  ] = \theta_k (1) $. Define now
\begin{eqnarray}
\bm{U}_n =     \left [
    \begin{array}{c}
      U_{n,0}  \\
      U_{n,1}
      \end{array}
    \right] . \nonumber
\end{eqnarray}
From (\ref{eq:stat-u_1}) it is seen that
\begin{eqnarray}
\sqrt{n} ( \widehat{\bm{\theta}}^{HT}  - \bm{\theta} ) = \sqrt{n} ( \bm{U}_n -  \bm{\theta} (1) ) + o_P (1)
\label{eq:stat-u_3}
\end{eqnarray}
Finally, taking $h^*_k ( Y_i , \, T_i , \, \bm{X}_i  ) = (  E [ h_k ( Y_i , \, T_i , \, \bm{X}_i ; \, Y_j , \, T_j , \, \bm{X}_j ) \vert Y_i , \, T_i , \, \bm{X}_i ] - \theta_k (1) )/2 $,
assuming that $ V ( h^*_k ( Y_i , \, T_i , \, \bm{X}_i  ) >0$ for $k=0, \, 1$, from standard theory of $U$-statistics (cfr. \cite{serfling80}, p. 192),
it is easy to argue that $\sqrt{n} ( \bm{U}_n  -\bm{\theta} (1) ) $ possesses the same limiting distribution as
\begin{eqnarray}
\frac{1}{\sqrt{n}} \sum_{i=1}^{n}  \bm{h}^* ( Y_i , \, T_i , \, \bm{X}_i  )
\nonumber
\end{eqnarray}
\noindent that turns out to be normal with zero expectation and variance-covariance matrix $\bm{\Sigma}$.
This shows statements $1$ and $3$. Statement $2$ can be proved by a technique similar to that of Proposition
$\ref{prop_consist}$.
\end{proof}

\begin{Lemma}
\label{lemma_3}
Suppose that $p_k ( \bm{x} ; \, \bm{\beta} )$ is a continuous function of $\bm{\beta}$ for each fixed $\bm{x}$, that assumptions $A1$-$A3$ in \cite{white82} and $A1$, $A2$ are satisfied, and that there exists $\delta >0$ such that $\delta \leq
p_k ( \bm{x} ; \, \bm{\beta}^* ) \leq 1- \delta$ for each $\bm{x}$. The following two statements hold.
\begin{eqnarray}
\sup_{\mathbf{x}} \left \vert  p_k ( \bm{x} ; \, \widehat{\bm{\beta}}_n ) -
p_k ( \bm{x} ; \, \bm{\beta}^* )  \right \vert & \stackrel{p}{\rightarrow} & 0 \;\;
{\mathrm{as}} \; n \rightarrow \infty \;\; \forall \, k=0, \, 1; \nonumber \\
\sup_{\bm{x}} \left \vert  \frac{1}{p_k ( \bm{x} ; \, \widehat{\bm{\beta}}_n )} -
\frac{1}{p_k ( \bm{x} ; \, \bm{\beta}^* )}
 \right \vert & \stackrel{p}{\rightarrow} & 0 \;\;
{\mathrm{as}} \; n \rightarrow \infty \;\; \forall \, k=0, \, 1  . \nonumber
\end{eqnarray}
\end{Lemma}
\begin{proof} Similar to the proof of Lemma $\ref{lemma_1}$.
\end{proof}

\begin{proof}[\textbf{Proof of Proposition \ref{HT_H_asympt_misspec}}]
Similarly to Proposition $\ref{prop_consist}$ consider the pseudo-estimator
\begin{eqnarray}
\widetilde{\theta}^{HT}_k = \frac{1}{n} \sum_{i=1}^{n}
 \frac{I_{(T_{i} =k)}}{p_k ( \bm{X}_{i} ; \,
\bm{\beta}^{*} )} I_{(Y_{i} =1)} , \;\; k=0, \, 1.
\nonumber
\end{eqnarray}
By repeating {\em verbatim} the arguments of Proposition $\ref{prop_consist}$, it is not difficult to see that
\begin{eqnarray}
\left \vert  \widehat{\theta}^{HT}_k = - \widetilde{\theta}^{HT}_k \right \vert \leq
\sup_{\bm{x}} \left \vert  \frac{1}{p_k ( \bm{x} ; \, \widehat{\bm{\beta}}_n )} -
\frac{1}{p_k ( \bm{x} ; \, \bm{\beta}^* )}
 \right \vert \stackrel{p}{\rightarrow} 0 \label{eq:conv_1a}
\end{eqnarray}
\noindent as $n \rightarrow \infty$, in view of Lemma $\ref{lemma_3}$. In the second place, from the Weak Law of Large Numbers, we also have
\begin{eqnarray}
\widetilde{\theta}^{HT}_k & \stackrel{p}{\rightarrow} & E \left [
\frac{1}{p_k ( \bm{X} ; \, \bm{\beta}^* )} I_{(Y =1)} I_{(T =k)}
\right ] \nonumber \\
& = & E \left [ \theta_k (1 \vert \bm{X} ) \frac{p_{k} ( \bm{X} )}{p_k ( \bm{X} ; \, \bm{\beta}^* )} \right ] \nonumber
\end{eqnarray}
\noindent as $n$ increases, and hence, from (\ref{eq:conv_1a}),
\begin{eqnarray}
\widehat{\theta}^{HT}_k \stackrel{p}{\rightarrow}  E \left [ \theta_k (1 \vert X ) \frac{p_{k} ( \bm{X} )}{p_k ( \bm{X} ; \, \bm{\beta}^* )} \right ] \;\; {\mathrm{as}} \; n \rightarrow \infty . \nonumber
\end{eqnarray}
Taking now into account that $E [ \theta_k ( 1 \vert \bm{X} ) ] = \theta_k$, relationship  (\ref{eq_ht_misspec}) is obtained.

As far as $\widehat{\theta}^{H}_k$ is concerned, using the above arguments it is immediate to see that
\begin{eqnarray}
\frac{1}{n} \sum_{i=1}^{n} \frac{1}{p_k ( \bm{X}_{i} ; \, \bm{\beta}^* )} \stackrel{p}{\rightarrow}
 E \left [ \frac{p_{k} ( \bm{X} )}{p_k ( \bm{X} ; \, \bm{\beta}^* )} \right ]
\nonumber
\end{eqnarray}
\noindent and hence, from (\ref{eq_ht_misspec} ,
\begin{eqnarray}
\widehat{\theta}^{H}_k \stackrel{p}{\rightarrow} \frac{E \left [ \theta_k (1 \vert X ) \frac{p_{k} ( \bm{X} )}{p_k
( \bm{X} ; \, \bm{\beta}^* )} \right ]}{E \left [ \frac{p_{k} ( \bm{X} )}{p_k ( \bm{X} ; \, \bm{\beta}^* )} \right ]} \;\;
{\mathrm{as}} \; n \rightarrow \infty   \nonumber
\end{eqnarray}
\noindent which is equivalent to (\ref{eq_h_misspec}).
\end{proof}




\end{appendices}


\input {sn-article.bbl}

\end{document}

%% file: sn-article.bbl

%% file: Vicard.bbl
\begin{thebibliography}{27}
\ifx \bisbn   \undefined \def \bisbn  #1{ISBN #1}\fi
\ifx \binits  \undefined \def \binits#1{#1}\fi
\ifx \bauthor  \undefined \def \bauthor#1{#1}\fi
\ifx \batitle  \undefined \def \batitle#1{#1}\fi
\ifx \bjtitle  \undefined \def \bjtitle#1{#1}\fi
\ifx \bvolume  \undefined \def \bvolume#1{\textbf{#1}}\fi
\ifx \byear  \undefined \def \byear#1{#1}\fi
\ifx \bissue  \undefined \def \bissue#1{#1}\fi
\ifx \bfpage  \undefined \def \bfpage#1{#1}\fi
\ifx \blpage  \undefined \def \blpage #1{#1}\fi
\ifx \burl  \undefined \def \burl#1{\textsf{#1}}\fi
\ifx \doiurl  \undefined \def \doiurl#1{\url{https://doi.org/#1}}\fi
\ifx \betal  \undefined \def \betal{\textit{et al.}}\fi
\ifx \binstitute  \undefined \def \binstitute#1{#1}\fi
\ifx \binstitutionaled  \undefined \def \binstitutionaled#1{#1}\fi
\ifx \bctitle  \undefined \def \bctitle#1{#1}\fi
\ifx \beditor  \undefined \def \beditor#1{#1}\fi
\ifx \bpublisher  \undefined \def \bpublisher#1{#1}\fi
\ifx \bbtitle  \undefined \def \bbtitle#1{#1}\fi
\ifx \bedition  \undefined \def \bedition#1{#1}\fi
\ifx \bseriesno  \undefined \def \bseriesno#1{#1}\fi
\ifx \blocation  \undefined \def \blocation#1{#1}\fi
\ifx \bsertitle  \undefined \def \bsertitle#1{#1}\fi
\ifx \bsnm \undefined \def \bsnm#1{#1}\fi
\ifx \bsuffix \undefined \def \bsuffix#1{#1}\fi
\ifx \bparticle \undefined \def \bparticle#1{#1}\fi
\ifx \barticle \undefined \def \barticle#1{#1}\fi
\bibcommenthead
\ifx \bconfdate \undefined \def \bconfdate #1{#1}\fi
\ifx \botherref \undefined \def \botherref #1{#1}\fi
\ifx \url \undefined \def \url#1{\textsf{#1}}\fi
\ifx \bchapter \undefined \def \bchapter#1{#1}\fi
\ifx \bbook \undefined \def \bbook#1{#1}\fi
\ifx \bcomment \undefined \def \bcomment#1{#1}\fi
\ifx \oauthor \undefined \def \oauthor#1{#1}\fi
\ifx \citeauthoryear \undefined \def \citeauthoryear#1{#1}\fi
\ifx \endbibitem  \undefined \def \endbibitem {}\fi
\ifx \bconflocation  \undefined \def \bconflocation#1{#1}\fi
\ifx \arxivurl  \undefined \def \arxivurl#1{\textsf{#1}}\fi
\csname PreBibitemsHook\endcsname

\bibitem[\protect\citeauthoryear{Conti and {De Giovanni}}{2022}]{condg22}
\begin{botherref}
\oauthor{\bsnm{Conti}, \binits{P.L.}},
\oauthor{\bsnm{{De Giovanni}}, \binits{L.}}:
Testing for the presence of treatment effect under selection on observables.
AStA Advances in Statistical Analysis,
10--100710182022004548
(2022)
\end{botherref}
\endbibitem

\bibitem[\protect\citeauthoryear{Donald and Hsu}{2014}]{donhsu14}
\begin{barticle}
\bauthor{\bsnm{Donald}, \binits{S.G.}},
\bauthor{\bsnm{Hsu}, \binits{Y.C.}}:
\batitle{Estimation and inference for distribution functions and quantile
  functions in treatment effect models}.
\bjtitle{Journal of Econometrics}
\bvolume{178},
\bfpage{383}--\blpage{397}
(\byear{2014})
\end{barticle}
\endbibitem

\bibitem[\protect\citeauthoryear{Abadie and Imbens}{2016}]{abadieImbens:16}
\begin{barticle}
\bauthor{\bsnm{Abadie}, \binits{A.}},
\bauthor{\bsnm{Imbens}, \binits{G.W.}}:
\batitle{Matching on the estimated propensity score}.
\bjtitle{Econometrica}
\bvolume{84},
\bfpage{781}--\blpage{807}
(\byear{2016})
\end{barticle}
\endbibitem

\bibitem[\protect\citeauthoryear{Ding}{2017}]{ding17}
\begin{barticle}
\bauthor{\bsnm{Ding}, \binits{P.}}:
\batitle{A {P}aradox from {Randomization}-{B}ased {C}ausal {I}nference}.
\bjtitle{Statistical Science}
\bvolume{32},
\bfpage{331}--\blpage{345}
(\byear{2017})
\end{barticle}
\endbibitem

\bibitem[\protect\citeauthoryear{Wu and Ding}{2018}]{wuding18}
\begin{botherref}
\oauthor{\bsnm{Wu}, \binits{J.}},
\oauthor{\bsnm{Ding}, \binits{P.}}:
Randomization tests for weak null hypotheses in randomized experiments.
\rm{arXiv:1809.07419 [stat.ME]}
(2018)
\end{botherref}
\endbibitem

\bibitem[\protect\citeauthoryear{Imbens and Rubin}{2015}]{imbrub:15}
\begin{bbook}
\bauthor{\bsnm{Imbens}, \binits{G.W.}},
\bauthor{\bsnm{Rubin}, \binits{D.B.}}:
\bbtitle{Causal Inference for Statistics, Social, and Biomedical Sciences}.
\bpublisher{Cambridge University Press, Cambridge}, \blocation{???}
(\byear{2015})
\end{bbook}
\endbibitem

\bibitem[\protect\citeauthoryear{Imbens and Wooldridge}{2009}]{imbenswool09}
\begin{barticle}
\bauthor{\bsnm{Imbens}, \binits{G.W.}},
\bauthor{\bsnm{Wooldridge}, \binits{J.M.}}:
\batitle{Recent developments in the econometrics of program evaluation}.
\bjtitle{Journal of Economic Literature}
\bvolume{47},
\bfpage{5}--\blpage{86}
(\byear{2009})
\end{barticle}
\endbibitem

\bibitem[\protect\citeauthoryear{Abadie}{2002}]{abadie02}
\begin{barticle}
\bauthor{\bsnm{Abadie}, \binits{A.}}:
\batitle{Bootstrap {T}ests for {D}istributional {T}reatment {E}ffects in
  {I}nstrumental {V}ariable {M}odels}.
\bjtitle{Journal of the American Statistical Association}
\bvolume{97},
\bfpage{284}--\blpage{292}
(\byear{2002})
\end{barticle}
\endbibitem

\bibitem[\protect\citeauthoryear{Lu et~al.}{2019}]{luzhangding19}
\begin{botherref}
\oauthor{\bsnm{Lu}, \binits{J.}},
\oauthor{\bsnm{Zhang}, \binits{Y.}},
\oauthor{\bsnm{Ding}, \binits{P.}}:
Sharp bounds on the relative treatment effect for ordinal outcomes.
Biometrics
\textbf{doi: 10.1111/biom.13148}
(2019)
\end{botherref}
\endbibitem

\bibitem[\protect\citeauthoryear{Cowell et~al.}{1999}]{cowell99}
\begin{bbook}
\bauthor{\bsnm{Cowell}, \binits{R.G.}},
\bauthor{\bsnm{Dawid}, \binits{A.P.}},
\bauthor{\bsnm{Lauritzen}, \binits{S.L.}},
\bauthor{\bsnm{Spiegelhalter}, \binits{D.J.}}:
\bbtitle{Probabilistic Networks and Expert Systems}.
\bpublisher{Springer},
\blocation{New York}
(\byear{1999})
\end{bbook}
\endbibitem

\bibitem[\protect\citeauthoryear{Lauritzen}{1996}]{lauritzen96}
\begin{bbook}
\bauthor{\bsnm{Lauritzen}, \binits{S.L.}}:
\bbtitle{Graphical {M}odels}.
\bpublisher{=xford University Press},
\blocation{Oxford}
(\byear{1996})
\end{bbook}
\endbibitem

\bibitem[\protect\citeauthoryear{Drton and Maathuis}{2017}]{drtmal17}
\begin{barticle}
\bauthor{\bsnm{Drton}, \binits{M.}},
\bauthor{\bsnm{Maathuis}, \binits{M.H.}}:
\batitle{Structure learning in graphical modeling}.
\bjtitle{Annual Review of Statistics and Its Application}
\bvolume{4},
\bfpage{365}--\blpage{393}
(\byear{2017})
\end{barticle}
\endbibitem

\bibitem[\protect\citeauthoryear{Cam}{1953}]{lecam53}
\begin{barticle}
\bauthor{\bsnm{Cam}, \binits{L.L.}}:
\batitle{On some asymptotic properties of maximum likelihood estimates and
  related {B}ayes estimates}.
\bjtitle{Univ. California Publ. Statist.}
\bvolume{1},
\bfpage{277}--\blpage{330}
(\byear{1953})
\end{barticle}
\endbibitem

\bibitem[\protect\citeauthoryear{Cam}{1960}]{lecam60}
\begin{barticle}
\bauthor{\bsnm{Cam}, \binits{L.L.}}:
\batitle{Locally asymptotically normal families of distributions}.
\bjtitle{Univ. California Publ. Statist.}
\bvolume{3},
\bfpage{27}--\blpage{98}
(\byear{1960})
\end{barticle}
\endbibitem

\bibitem[\protect\citeauthoryear{Balov}{2013}]{balov13}
\begin{barticle}
\bauthor{\bsnm{Balov}, \binits{N.}}:
\batitle{Consistent model selection of discrete {B}ayesian networks from
  incomplete data}.
\bjtitle{Electronic Journal of Statistics}
\bvolume{7},
\bfpage{1935}--\blpage{7524}
(\byear{2013})
\end{barticle}
\endbibitem

\bibitem[\protect\citeauthoryear{Nandy et~al.}{2018}]{nandy18}
\begin{barticle}
\bauthor{\bsnm{Nandy}, \binits{P.}},
\bauthor{\bsnm{Hauser}, \binits{A.}},
\bauthor{\bsnm{Maathuis}, \binits{M.H.}}:
\batitle{High-{D}imensional {C}onsistency in {S}core-{B}ased and {H}ybrid
  {S}tructure {L}earning}.
\bjtitle{The Annals of Statistics}
\bvolume{46},
\bfpage{3151}--\blpage{3183}
(\byear{2018})
\end{barticle}
\endbibitem

\bibitem[\protect\citeauthoryear{Hahn}{1998}]{hahn98}
\begin{barticle}
\bauthor{\bsnm{Hahn}, \binits{J.}}:
\batitle{On the {R}ole of the {P}ropensity {S}core in {E}fficient
  {S}emiparametric {E}stimation of {A}verage {T}reatment {E}ffects}.
\bjtitle{Econometrica}
\bvolume{66},
\bfpage{315}--\blpage{331}
(\byear{1998})
\end{barticle}
\endbibitem

\bibitem[\protect\citeauthoryear{Zubizzarreta}{2015}]{zubi15}
\begin{barticle}
\bauthor{\bsnm{Zubizzarreta}, \binits{J.R.}}:
\batitle{Stable weights that balance covariates for estimation with incomplete
  outcome data}.
\bjtitle{Journal of the American Statistical Association}
\bvolume{110},
\bfpage{910}--\blpage{922}
(\byear{2015})
\end{barticle}
\endbibitem

\bibitem[\protect\citeauthoryear{Chattopadhyay et~al.}{2020}]{chatto20}
\begin{barticle}
\bauthor{\bsnm{Chattopadhyay}, \binits{A.}},
\bauthor{\bsnm{Hase}, \binits{C.H.}},
\bauthor{\bsnm{Zubizarreta}, \binits{J.R.}}:
\batitle{Balancing vs modeling approaches to weighting in practice}.
\bjtitle{Statistics in Medicine}
\bvolume{39},
\bfpage{3227}--\blpage{3254}
(\byear{2020})
\end{barticle}
\endbibitem

\bibitem[\protect\citeauthoryear{Ben-Michael et~al.}{2021}]{benmi21}
\begin{botherref}
\oauthor{\bsnm{Ben-Michael}, \binits{E.}},
\oauthor{\bsnm{Feller}, \binits{A.}},
\oauthor{\bsnm{Hirshberg}, \binits{D.A.}},
\oauthor{\bsnm{Zubizarreta}, \binits{J.R.}}:
The balancing act in causal inference.
arXiv preprint arXiv:2110.14831
(2021)
\end{botherref}
\endbibitem

\bibitem[\protect\citeauthoryear{Hern\'{a}n and Robins}{2006}]{hernrob06}
\begin{barticle}
\bauthor{\bsnm{Hern\'{a}n}, \binits{M.A.}},
\bauthor{\bsnm{Robins}, \binits{J.M.}}:
\batitle{Estimating causal effects from epidemiological data}.
\bjtitle{Journal of Epidemiology and Community Health}
\bvolume{60},
\bfpage{578}--\blpage{586}
(\byear{2006})
\end{barticle}
\endbibitem

\bibitem[\protect\citeauthoryear{Lunceford and Davidian}{2004}]{luncdav04}
\begin{barticle}
\bauthor{\bsnm{Lunceford}, \binits{J.K.}},
\bauthor{\bsnm{Davidian}, \binits{M.}}:
\batitle{tratification and weighting via the propensity score in estimation of
  causal treatment effects: a comparative study}.
\bjtitle{Statistics in Medicine}
\bvolume{23},
\bfpage{2937}--\blpage{2960}
(\byear{2004})
\end{barticle}
\endbibitem

\bibitem[\protect\citeauthoryear{Kim}{2019}]{kim19}
\begin{barticle}
\bauthor{\bsnm{Kim}, \binits{K.I.}}:
\batitle{Efficiency of {A}verage {T}reatment {E}ffect {E}stimation {W}hen the
  {T}rue {P}ropensity {I}s {P}arametric}.
\bjtitle{Econometrics https://doi.org/10.3390/econometrics7020025}
\bvolume{7},
\bfpage{2}--\blpage{25}
(\byear{2019})
\end{barticle}
\endbibitem

\bibitem[\protect\citeauthoryear{Sen}{1960}]{sen60}
\begin{barticle}
\bauthor{\bsnm{Sen}, \binits{P.K.}}:
\batitle{On some convergence properties of $u$-statistics}.
\bjtitle{Calcutta Statistical Association Bulletin}
\bvolume{10},
\bfpage{1}--\blpage{18}
(\byear{1960})
\end{barticle}
\endbibitem

\bibitem[\protect\citeauthoryear{Hirano et~al.}{2003}]{hirano03}
\begin{barticle}
\bauthor{\bsnm{Hirano}, \binits{K.}},
\bauthor{\bsnm{Imbens}, \binits{G.W.}},
\bauthor{\bsnm{Ridder}, \binits{G.}}:
\batitle{Efficient {E}stimation of {A}verage {T}reatment {E}ffects {U}sing the
  {E}stimated {P}ropensity {S}core}.
\bjtitle{Econometrica}
\bvolume{71},
\bfpage{1161}--\blpage{1189}
(\byear{2003})
\end{barticle}
\endbibitem

\bibitem[\protect\citeauthoryear{White}{1982}]{white82}
\begin{barticle}
\bauthor{\bsnm{White}, \binits{H.}}:
\batitle{Maximum {L}ikelihood {E}stimation of {M}isspecified {M}odels}.
\bjtitle{Econometrica}
\bvolume{50},
\bfpage{1}--\blpage{25}
(\byear{1982})
\end{barticle}
\endbibitem

\bibitem[\protect\citeauthoryear{Serfling}{1980}]{serfling80}
\begin{bbook}
\bauthor{\bsnm{Serfling}, \binits{R.J.}}:
\bbtitle{Approximation {T}heorems of {M}athematical {S}tatistics}.
\bpublisher{Wiley},
\blocation{New York}
(\byear{1980})
\end{bbook}
\endbibitem

\end{thebibliography}
